\documentclass[rmp,amsfonts,showpacs,showkeys,preprint]{revtex4}
\usepackage{xcolor}
\usepackage{graphicx}
\usepackage{dcolumn}
\usepackage{bm}
\usepackage{eepic}
\bibpunct{[}{]}{,}{a}{}{;}
\RequirePackage{times}
\RequirePackage{mathptm}
\usepackage{longtable}

\begin{document}

\title{A glance at singlet states and four-partite correlations}

\author{Maria Schimpf}
\author{Karl Svozil}
\email{svozil@tuwien.ac.at}
\homepage{http://tph.tuwien.ac.at/~svozil}
\affiliation{Institute for Theoretical Physics, Vienna University of Technology,  \\
Wiedner Hauptstra\ss e 8-10/136, A-1040 Vienna, Austria}

\begin{abstract}
Group theoretic methods to construct all $N$-particle singlet states by iterative recursion are presented.
These techniques are applied to the quantum correlations of four spin one-half particles in their singlet states.
Multipartite quantized systems can be partitioned, and their observables grouped and redefined into condensed correlations.
\end{abstract}


\maketitle

\tableofcontents
\newpage

\section{Introduction}

Improved experimental particle production techniques
and potential applications in quantum information theory
have stimulated interest in
multipartite singlet and other entangled states.
In particular, singlet states are among the most useful states in quantum
mechanics, as they appear form-invariant under spatial rotations.
Hence, a physical property such as uniqueness~\cite{svozil-2006-uniquenessprinciple}
or equibalance~\cite{zeil-99}
which holds true in one frame or direction
remains to be true in all other frames or directions obtained by spatial rotations.

Yet, the explicit structure of singlet states  ---  although well understood
in general terms in group theory  ---  has up to now neither been
enumerated nor investigated beyond a few instances for
spin one-half and spin one particles. Recent theoretical and
experimental studies in multi-particle production (e.g.,
Ref.~\cite{egbkzw}) suggest that a more systematic way to generate
the complete set of arbitrary $N$-particle singlet states is desirable.

In the present study we first pursue an algorithmic generation strategy,
and tabulate some of the first singlet states. The recursive
method employed is based on triangle relations and
Clebsch-Gordan coefficients (e.g., Ch.~13, Sec.~27 of
Ref.~\cite{messiah-62}). With this approach, a complete table of
all angular momentum states can be enumerated.
The singlet states are obtained {\it via}  the various pathways towards the $j = m = 0$ states.

The
procedure can best be illustrated by a triangular diagram, where
the states in ascending order of angular momentum are drawn against the number
of particles. In such a diagram, the ``lowest'' states correspond
to singlet states.

Consider, for the sake of an explicit demonstration of this generation method, a two-dimensional diagram such as the ones depicted in Fig.~\ref{2005-singlet-f12-e1}
and in Fig.~\ref{2005-singlet-f1-e2},
which represents the ``space'' or ``domain'' of all multi-partite states.
In such a diagram, the {\em number of particles} is represented by the abscissa (the $x$-coordinate) along the positive $x$-axis.
The ordinate (the $y$-coordinate) of the state is equal the total angular momentum of the state.
Note that a single point may represent many states; all corresponding to an equal number of particles,
and all having the same total angular momentum.
$N$-partite singlet states can be constructed by starting from the unique state of one particle,
then proceeding {\it via} all ``diagonal'' and, whenever possible for integer spins,
also ``horizontal'' pathways (e.g., the ``horizontal'' path in Fig.~\ref{2005-singlet-f1-e2})
consisting of single substeps adding one particle after the other
--- either diagonally from the lower left to the upper right ``{\color{blue}$\nearrow$},''
or diagonally from the upper left to the lower right ``{\color{blue}$\searrow$},''
or, if possible, also horizontally from left to right ``{\color{blue}$\rightarrow$}'' ---
towards the zero momentum state of $N$ particles.
Every diagonal or horizontal substep corresponds to the addition of a single particle.
Below we shall explicitly construct singlet states composed from particles of spin one-half and spin one.

In the second part of this article, we present an explicit analysis of the singlet states of
four spin one-half particles in terms of their probabilities and
expectation functions for spin state measurements.
We also investigate the possibility to group the outcomes
of the four spin state measurements on each particle to obtain ``condensed''
observables. Likewise, we consider selection of one or two particles and the resulting
correlations.
One of our physical motivations for doing so was the question of how such ``condensed'' observables
would perform with respect to violations of classical locality conditions.

\section{General algorithm for obtaining singlet states}

In what follows we present a method to construct all states for a given
number of particles. They are the basis to construct non-trivial,
e.g., non-``zigzag''  singlet states, which are not just
products of singlet states of a smaller number of particles.
Although only the spin one-half and the spin one cases are
explicitly discussed, the method applies to arbitrary spin.

\subsection{Spin one-half}

We start by considering the spin state of a single spin one-half particle. A second spin one-half particle is added by
combining two angular momenta $\frac{1}{2}$ to all possible total
angular momenta $l=0$ or $1$. Next, a third particle is introduced
by coupling a third angular momentum $\frac{1}{2}$ to all
previously derived states. Following the triangle equation, the
resulting $j$-values for each $l$ are in the domain
\begin{equation}
|l-j_3|\leq  j\leq l+j_3.
\end{equation}

In order to obtain all $N$-particle singlet states, we
successively produce all states (not only singlet states) of
$N/2$ particles. Note that for $N/2\leq
h\leq N$ particles we only need angular momentum states within
$0\leq j\leq (N-h)/2$, because the construction method does not allow states with higher angular momentum to
``bend diagonally backwards'' and finally reach the angular momentum zero singlet state.

Angular momentum states will be written as
$|h,j,m,i\rangle$, where $h$ denotes the particle number, $j$ the
angular momentum, and $m$ the magnetic quantum number.
Note that there may exist many states with equal $h$, $j$ and $m$.
Thus $i$ denotes the number
of state in an enumeration of all $h$-partite states with identical angular momentum $j$ and magnetic quantum number $m$.
In the enumeration scheme chosen, we first take states generated from higher total angular momentum,
followed by states with equal total angular momentum for spin one particles, and states with lower  total angular momentum.
For spin one-half particles, let us define a function $f(j+1,h)$ denoting the {\em total number} of
states of $h$ particles with total angular momentum $j/2$.
This function is
tabulated in Table~\ref{2008-sg-tsoh-numbers}
for the spin one-half particle case.
The Clebsch-Gordan coefficient is denoted by
$\langle j_1,j_2,m_1,m_2|j,m\rangle$.

For spin one-half particles, an arbitrary state $|h,j,m,i\rangle$, $h>1$, can be generated
from the states with one particle less by {\em adding} a particle,
thereby {\em increasing} or {\em decreasing} the total angular momentum of the previous state containing one particle less.
Thus, we obtain two different pathways towards $|h,j,m,i\rangle$;
one from the total angular momentum $j+{1\over 2}$, symbolized graphically by ``{\color{blue}$\searrow$},''
and one from the total angular momentum $j-{1\over 2}$, symbolized graphically by ``{\color{blue}$\nearrow$}.''

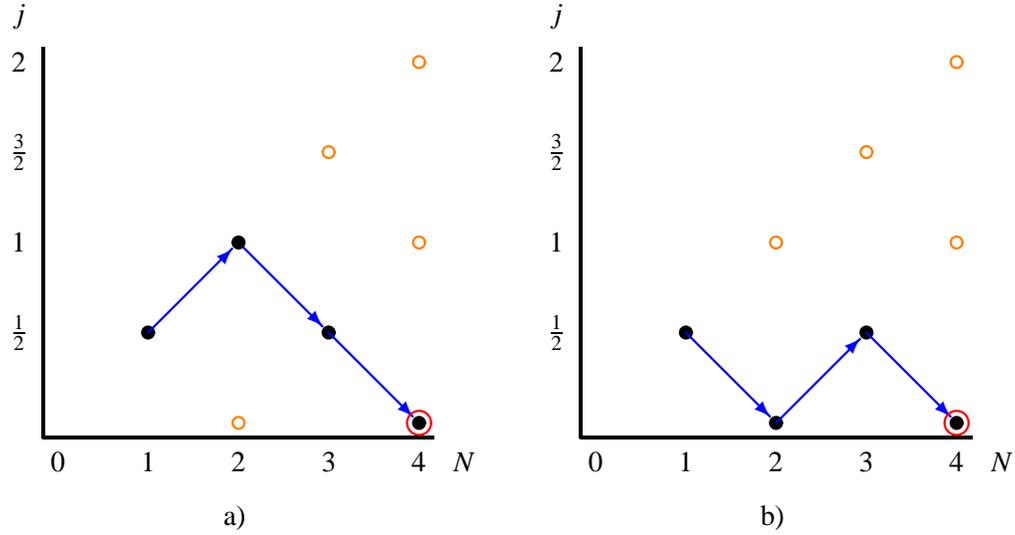
\begin{figure}
\begin{center}
\begin{tabular}{ccc}
\unitlength 0.40mm
\allinethickness{1pt} 
\begin{picture}(150.00,150.00)
\put(15.00,2.00){\makebox(0,0)[cc]{$0$}}
\put(45.00,2.00){\makebox(0,0)[cc]{$1$}}
\put(75.00,2.00){\makebox(0,0)[cc]{$2$}}
\put(105.0,2.00){\makebox(0,0)[cc]{$3$}}
\put(135.00,2.00){\makebox(0,0)[cc]{$4$}}
\put(150.00,2.00){\makebox(0,0)[cc]{$N$}}
\put(2.00,45.00){\makebox(0,0)[cc]{${1\over 2}$}}
\put(2.00,75.00){\makebox(0,0)[cc]{$1$}}
\put(2.00,105.00){\makebox(0,0)[cc]{${3\over 2}$}}
\put(2.00,135.00){\makebox(0,0)[cc]{$2$}}
\put(2.00,150.00){\makebox(0,0)[cc]{$j$}}
\put(10.00,10.00){\line(0,1){130.00}}
\put(10.00,10.00){\line(1,0){130.00}}
\put(45.00,45.00){\color{blue} \circle*{4.00}} \put(75.00,15.00){\color{orange} \circle{4.00}}
\put(105.00,45.00){\color{blue} \circle*{4.00}}
\put(135.00,15.00){\color{blue} \circle*{4.00}}
\put(75.00,75.00){\color{blue} \vector(1,-1){28.00}}
\put(45.00,45.00){\color{blue} \vector(1,1){28.00}}
\put(105.00,45.00){\color{blue} \vector(1,-1){28.00}}
\put(75.00,75.00){\color{blue} \circle*{4.00}}
\put(135.00,75.00){\color{orange} \circle{4.00}}
\put(105.00,105.00){\color{orange} \circle{4.00}}
\put(135.00,135.00){\color{orange} \circle{4.00}}
\put(135.00,15.00){\color{red} \circle{8.00}}
\end{picture}
&
$\qquad$
&
\unitlength 0.40mm
\allinethickness{1pt} 
\begin{picture}(150.00,150.00)
\put(15.00,2.00){\makebox(0,0)[cc]{$0$}}
\put(45.00,2.00){\makebox(0,0)[cc]{$1$}}
\put(75.00,2.00){\makebox(0,0)[cc]{$2$}}
\put(105.0,2.00){\makebox(0,0)[cc]{$3$}}
\put(135.00,2.00){\makebox(0,0)[cc]{$4$}}
\put(150.00,2.00){\makebox(0,0)[cc]{$N$}}
\put(2.00,45.00){\makebox(0,0)[cc]{${1\over 2}$}}
\put(2.00,75.00){\makebox(0,0)[cc]{$1$}}
\put(2.00,105.00){\makebox(0,0)[cc]{${3\over 2}$}}
\put(2.00,135.00){\makebox(0,0)[cc]{$2$}}
\put(2.00,150.00){\makebox(0,0)[cc]{$j$}}
\put(10.00,10.00){\line(0,1){130.00}}
\put(10.00,10.00){\line(1,0){130.00}}
\put(45.00,45.00){\color{blue} \circle*{4.00}}
\put(75.00,15.00){\color{blue} \circle*{4.00}}
\put(105.00,45.00){\color{blue} \circle*{4.00}}
\put(135.00,15.00){\color{blue} \circle*{4.00}}
\put(45.00,45.00){\color{blue} \vector(1,-1){28.00}}
\put(75.00,15.00){\color{blue} \vector(1,1){28.00}}
\put(105.00,45.00){\color{blue} \vector(1,-1){28.00}}
\put(75.00,75.00){\color{orange} \circle{4.00}}
\put(135.00,75.00){\color{orange} \circle{4.00}}
\put(105.00,105.00){\color{orange} \circle{4.00}}
\put(135.00,135.00){\color{orange} \circle{4.00}}
\put(135.00,15.00){\color{red} \circle{8.00}}
\end{picture}
\\
a)&&b)
\end{tabular}
\end{center}
\caption{Construction of both singlet states a) and b) of four
spin one-half particles. Concentric circles indicate the
target states. The second state is a ``zigzag'' state composed by the product of two two-partite singlet states.  \label{2005-singlet-f12-e1}}
\end{figure}

For the sake of demonstration of the method employed,
we shall explicitly discuss one of the two cases, in which the {\em addition} of one particle $h-1 \rightarrow h$
results in a {\em lowering} of the total angular momentum
by ${1\over 2}$ through $j+{1\over 2} \rightarrow j$,
thus representing the diagonal pathway ``{\color{blue}$\searrow$}'' from the ``upper left'' to the ``lower right'' in a
diagram (nonuniquely) representing states as points with coordinates given by the number of particles and the total angular momentum, respectively
(e.g., Fig.~\ref{2005-singlet-f12-e1}).
The first contribution,
associated with the magnetic quantum numbers $m-{1\over2}$ and $+{1\over2}$, can be constructed  from the
product state
\begin{equation}
|h-1,j+{1\over2},m-{1\over2},i\rangle  \otimes  |1,{1\over2},{1\over2},1\rangle
\end{equation}
by
multiplying it with the Clebsch-Gordan coefficient
\begin{equation}
\langle j+{1\over2},m-{1\over2},{1\over2},{1\over2}|j,m\rangle .
\end{equation}
Similarly, the second contribution  to $|h,j,m,i\rangle$,
associated with the magnetic quantum numbers $m+{1\over2}$ and $-{1\over2}$,
can be constructed {\it via}
the product state
\begin{equation}
|h-1,j+{1\over2},m+{1\over2},i\rangle\otimes|1,{1\over2},-{1\over2},1\rangle ,
\end{equation}
multiplied with the Clebsch-Gordan coefficient
\begin{equation}
\langle j+{1\over2},m+{1\over2},{1\over2},-{1\over2}|j,m\rangle .
\end{equation}
Adding
the two results, we obtain the state $|h,j,m,i\rangle$; i.e.,
\begin{equation}
\begin{array}{rcl}
|h,j,m,i\rangle &=&   \langle j+{1\over2},m-{1\over2},{1\over2},{1\over2}|j,m\rangle |h-1,j+{1\over2},m-{1\over2},i\rangle  \otimes  |1,{1\over2},{1\over2},1\rangle  +\\
&&\; + \langle j+{1\over2},m+{1\over2},{1\over2},-{1\over2}|j,m\rangle  |h-1,j+{1\over2},m+{1\over2},i\rangle\otimes|1,{1\over2},-{1\over2},1\rangle
.
\end{array}
\end{equation}

We do this
for $m= -j,\ldots,j$ and for all states labeled by the state number $i= 1,2,\ldots$, $f((2j+1)+1,h-1)$.
Recall that
$f(j+1,h)$ denotes the total number of
states of $h$ particles with angular momentum $j/2$.
It can be computed by counting the number of all states generated by all possible pathways in the construction method described above.

Similarly, if $j$ is greater than
zero, we obtain the
state  $|h,j,m,i\rangle$
from the diagonal pathway ``{\color{blue}$\nearrow$},''
starting from
the states $|h-1,j-{1\over2},m-{1\over2},i\rangle$ and
$|h-1,j-{1\over2},m+{1\over2},i\rangle$ of $h-1$ particles and total angular momentum $j-{1\over2}$
 by adding a single particle {\it via}
\begin{equation}\langle
j-{1\over2},m-{1\over2},{1\over2},{1\over2}|j,m\rangle|h-1,j-{1\over2},m-{1\over2},i\rangle\otimes|1,{1\over2},{1\over2},1\rangle
\end{equation}
and
\begin{equation}\langle
j-{1\over2},m+{1\over2},{1\over2},-{1\over2}|j,m\rangle|h-1,j-{1\over2},m+{1\over2},i\rangle\otimes|1,{1\over2},-{1\over2},1\rangle .
\end{equation}
This procedure is carried out for $m= -j,\ldots,j$ and $i$ satisfying
\begin{equation}
f((2j+1)+1,h-1)+1 \le i \le f((2j+1)+1,h-1)+f((2j+1)-1,h-1).
\end{equation}


A concrete example is drawn in Fig.~\ref{2005-singlet-f12-e1}.
It contains the pathways leading to the construction of both
singlet states of four spin one-half particles.

For spin one-half particles, the function $f(j+1,h)$ denoting the {\em total number} of
states of $h$ particles with total angular momentum $j/2$ is
tabulated in
Table~\ref{2008-sg-tsoh-numbers}.
The bottom line above the axis contains the  number of different orthogonal singlet states.
\begin{table}
\begin{center}
\begin{tabular}{c|cccccccccccccccccccccc}
\multicolumn{1}{c}{$j$}\\
5                     &   &  &  &  &  &  &  &  &  & 1 &  &  &  &  &  &  &  &  &  &  \\
$\frac{9}{2}$         &   &  &  &  &  &  &  &  & 1 &  & 10 &  &  &  &  &  &  &  &  &  \\
4                     &   &  &  &  &  &  &  & 1 &  & 9 &  & 54 &  &  &  &  &  &  &  &  \\
$\frac{7}{2}$         &   &  &  &  &  &  & 1 &  & 8 &  & 44 &  & 208 &  &  &  &  &  &  &  \\
3                     &   &  &  &  &  & 1 &  & 7 &  & 35 &  & 154 &  & 637 &  &  &  &  &  &  \\
$\frac{5}{2}$         &   &  &  &  & 1 &  & 6 &  & 27 &  & 110 &  & 429 &  & 1638 &  &  &  &  &  \\
2                     &   &  &  & 1 &  & 5 &  & 20 &  & 75 &  & 275 &  & 1001 &  & 3640 &  &  &  &  \\
$\frac{3}{2}$         &   &  & 1 &  & 4 &  & 14 &  & 48 &  & 165 &  & 572 &  & 2002 &  & 7072 &  &  &  \\
1                     & & 1 &  & 3 &  & 9 &  & 28 &  & 90 &  & 297 &  & 1001 &  & 3432 &  & 11934 &  &  \\
$\frac{1}{2}$         & 1 &  & 2 &  & 5 &  & 14 &  & 42 &  & 132 &  & 429 &  & 1430 &  & 4862 &  & 16796 &  \\
0                     & & 1 &  & 2 &  & 5 &  & 14 &  & 42 &  & 132 &  & 429 &  & 1430 &  & 4862 &  & 16796\\
\cline{2-22}
\multicolumn{1}{c}{ }&1&2&3&4&5&6&7&8&9&10&11&12&13&14&15&16&17&18&19&20 &$N$
\end{tabular}
\end{center}
\caption{\label{2008-sg-tsoh-numbers} Enumeration of the total numbers of states contributing to a calculation of singlet states up to 20 spin one-half particles.
The bottom line above the axis shows the actual number of different orthogonal singlet states.}
\end{table}
The singlet states of up to six spin one-half  particles are explicitly enumerated in Table \ref{2005-singlet-t12}.



\begin{table}
\begin{tabular}{ccc}
\hline\hline N & \# & \\\hline\hline

2&1&$\frac{1}{{\sqrt{2}}}\big(|+-\rangle-|-+\rangle\big);$\\\hline
4&1&$-\frac{1}{2\sqrt{3}} \big(|-+-+\rangle
+|-++-\rangle+|+--+\rangle +|+-+-\rangle \big)+$\\&&$
+\frac{1}{\sqrt{3}}\big(|--++\rangle +|++--\rangle\big)
;$\\

4&2&$\big(-\frac{1}{{\sqrt{2}}}|-+\rangle
+\frac{1}{{\sqrt{2}}}|+-\rangle\big)^2
;$\\\hline

6&1& $-\frac{1}{2}| --- +++\rangle +-\frac{1}{6}\big(
|-++--+\rangle + |-++-+-\rangle+$\\&&$
+|-+++--\rangle + |+-+--+\rangle +|+-+-+-\rangle
+$\\&&$+ |+-++--\rangle +|++ --- +\rangle +
|++--+-\rangle +$\\&&$+|++-+--\rangle\big) +
\frac{1}{6}\big(|--+-++\rangle +|--++-+\rangle +$\\&&$+
|--+++-\rangle +|-+--++\rangle + |-+-+-+\rangle
+$\\&&$+|-+-++-\rangle + |+ --- ++\rangle
+|+--+-+\rangle +$\\&&$+
|+--++-\rangle\big) +\frac{1}{2}|+++ --- \rangle;$\\

6&2&$ -\frac{{\sqrt{2}}}{3}|--+-++\rangle +-\frac{1}{3
{\sqrt{2}}}\big(|-+++--\rangle +|+-++--\rangle+$\\&&$
+|++ --- +\rangle+
 |++--+-\rangle\big) +-\frac{1}{6 {\sqrt{2}}}\big(|-+-+-+\rangle+$\\&&$ +|-+-++-\rangle
 +
 |+--+-+\rangle +|+--++-\rangle\big)+$\\&&$ +\frac{1}{6 {\sqrt{2}}}\big(|-++--+\rangle +
 |-++-+-\rangle +|+-+--+\rangle+$\\&&$ +|+-+-+-\rangle\big)+
 \frac{1}{3 {\sqrt{2}}}\big(|--++-+\rangle +|--+++-\rangle+$\\&&$ +|-+--++\rangle +
|+ --- ++\rangle\big)+\frac{{\sqrt{2}}}{3}|++-+--\rangle;$\\

6&3&$ -\frac{1}{{\sqrt{6}}}\big(|-+--++\rangle
+|-+++--\rangle\big)+-\frac{1}{2
{\sqrt{6}}}\big(|+--+-+\rangle +$\\&&$+ |+--++-\rangle+
 |+-+--+\rangle +|+-+-+-\rangle\big)+$\\&&$ +\frac{1}{2 {\sqrt{6}}}\big(|-+-+-+\rangle +
 |-+-++-\rangle +|-++--+\rangle
 +$\\&&$+|-++-+-\rangle\big)+
 \frac{1}{{\sqrt{6}}}\big(|+ --- ++\rangle +|+-++--\rangle\big)
 ;$\\

6&4&$ -\frac{1}{{\sqrt{6}}}\big(|--++-+\rangle +
|++ --- +\rangle\big) +-\frac{1}{2
{\sqrt{6}}}\big(|-+-++-\rangle+$\\&&$ +|-++-+-\rangle +
|+--++-\rangle +|+-+-+-\rangle\big)+$\\&&$ +\frac{1}{2
{\sqrt{6}}}\big(|-+-+-+\rangle +
 |-++--+\rangle +|+--+-+\rangle+$\\&&$ +|+-+--+\rangle\big) +
 \frac{1}{{\sqrt{6}}}\big(|--+++-\rangle
 +|++--+-\rangle\big);$\\

6&5&$\big(-\frac{1}{{\sqrt{2}}}|-+\rangle
+\frac{1}{{\sqrt{2}}}|+-\rangle\big)^3.$\\\hline\hline
\end{tabular}
\caption{First singlet states of $N$ spin one-half particles.
\label{2005-singlet-t12}}
\end{table}

\subsection{Spin one}

The construction of the singlet states of spin one particles follows similar rules as in the case of spin one-half particles.
One example is the construction of the  singlet state consisting of three  spin one particles drawn in
 Fig.~\ref{2005-singlet-f1-e2}.
Note that in this case, as for all particles of integer spin, there are {\em three} possible
subpaths per addition of one particle; two diagonal ``{\color{blue}$\nearrow$}'' and ``{\color{blue}$\searrow$}'' pathways as in the case for
spin one-half particles,
as well as one horizontal ``{\color{blue}$\rightarrow$}.''
\begin{figure}
\begin{center}
\unitlength 0.40mm
\allinethickness{1pt} 
\begin{picture}(120.00,120.00)
\put(15.00,2.00){\makebox(0,0)[cc]{$0$}}
\put(45.00,2.00){\makebox(0,0)[cc]{$1$}}
\put(75.00,2.00){\makebox(0,0)[cc]{$2$}}
\put(105.00,2.00){\makebox(0,0)[cc]{$3$}}
\put(120.00,2.00){\makebox(0,0)[cc]{$N$}}
\put(2.00,45.00){\makebox(0,0)[cc]{${1}$}}
\put(2.00,75.00){\makebox(0,0)[cc]{$2$}}
\put(2.00,105.00){\makebox(0,0)[cc]{${3}$}}
\put(2.00,120.00){\makebox(0,0)[cc]{$j$}}
\put(10.00,10.00){\line(0,1){100.00}}
\put(10.00,10.00){\line(1,0){100.00}}
\put(45.00,45.00){\color{blue} \circle*{4.00}}
\put(75.00,15.00){\color{orange} \circle{4.00}}
\put(105.00,15.00){\color{blue} \circle*{4.00}}
\put(105.00,15.00){\color{red} \circle{8.00}}
\put(75.00,44.67){\color{blue} \vector(1,-1){28.00}}
\put(75.00,75.00){\color{orange} \circle{4.00}}
\put(105.00,75.00){\color{orange} \circle{4.00}}
\put(105.00,105.00){\color{orange} \circle{4.00}}
\put(75.00,45.00){\color{blue} \circle*{4.00}}
\put(105.00,45.00){\color{orange} \circle{4.00}}
\put(45.00,45.00){\color{blue} \vector(1,0){27.00}}
\end{picture}
\end{center}
\caption{ Construction of the singlet state of three spin one
particles. Note that for integer spin, there are three possible
subpaths per addition of one particle; two diagonal ``{\color{blue}$\nearrow$}'' and ``{\color{blue}$\searrow$},''
as well as one horizontal ``{\color{blue}$\rightarrow$}.''
\label{2005-singlet-f1-e2}}
\end{figure}
Table~\ref{2008-sg-tso-numbers} enumerates the numbers of states contributing to a calculation of singlet states up to 18 spin one particles.
The bottom line above the axis shows the actual number of different orthogonal singlet states.
The singlet states of up to four spin one (with one singlet state of 5) particles are explicitly enumerated in Table \ref{2005-singlet-t1}.
\begin{table}
\begin{center}
\begin{tabular}{c|cccccccccccccccccccccccc}
\multicolumn{1}{c}{$j$}\\
9                   &    &  &  &  &  &  &  &  & 1 &  &  &  &  &  &  &  &  &  \\
8                   &    &  &  &  &  &  &  & 1 & 8 & 45 &  &  &  &  &  &  &  &  \\
7                   &    &  &  &  &  &  & 1 & 7 & 36 & 155 & 605 &  &  &  &  &  &  &  \\
6                   &    &  &  &  &  & 1 & 6 & 28 & 111 & 405 & 1397 & 4642 &  &  &  &  &  &  \\
5                   &    &  &  &  & 1 & 5 & 21 & 76 & 258 & 837 & 2640 & 8162 & 24882 &  &  &  &  &  \\
4                   &    &  &  & 1 & 4 & 15 & 49 & 154 & 468 & 1398 & 4125 & 12078 & 35178 & 102102 &  &  &  &  \\
3                   &    &  & 1 & 3 & 10 & 29 & 84 & 238 & 672 & 1890 & 5313 & 14938 & 42042 & 118482 & 334425 &  &  &  \\
2                   &    & 1 & 2 & 6 & 15 & 40 & 105 & 280 & 750 & 2025 & 5500 & 15026 & 41262 & 113841 & 315420 & 877320 &  &  \\
1                   &  1 & 1 & 3 & 6 & 15 & 36 & 91 & 232 & 603 & 1585 & 4213 & 11298 & 30537 & 83097 & 227475 & 625992 & 1730787 &  \\
0                   &    & 1 & 1 & 3 & 6 & 15 & 36 & 91 & 232 & 603 & 1585 & 4213 & 11298 & 30537 & 83097 & 227475 & 625992 & 1730787  \\
\cline{2-19}
\multicolumn{1}{c}{ }&1&2&3&4&5&6&7&8&9&10&11&12&13&14&15&16&17&18 &$N$
\end{tabular}
\end{center}
\caption{\label{2008-sg-tso-numbers} Enumeration of the total numbers of states contributing to a calculation of singlet states up to 18 spin one particles.
The bottom line above the axis shows the actual number of different orthogonal singlet states.}
\end{table}

\begin{table}
\begin{tabular}{ccc}
\hline\hline
N & \# & \\
\hline\hline
2&1&$\frac{1}{{\sqrt{3}}}\big(-|0,0\rangle+|-1,1\rangle+|1,-1\rangle\big);$\\\hline

3&1&$
-\frac{1}{{\sqrt{6}}}\big(|-1,0,1\rangle+|0,1,-1\rangle+|1,-1,0\rangle\big)+$\\&&$
+\frac{1}{{\sqrt{6}}}\big(|-1,1,0\rangle+|0,-1,1\rangle+|1,0,-1\rangle\big);$\\\hline

4&1&$ -\frac{1}{2 {\sqrt{5}}}\big(|-1,0,0,1\rangle+
|-1,0,1,0\rangle+|0,-1,0,1\rangle+
|0,-1,1,0\rangle+$\\&&$+|0,1,-1,0\rangle+
|0,1,0,-1\rangle+|1,0,-1,0\rangle+
|1,0,0,-1\rangle\big)+$\\&&$+\frac{1}{6
{\sqrt{5}}}\big(|-1,1,-1,1\rangle+
|-1,1,1,-1\rangle+|1,-1,-1,1\rangle+
|1,-1,1,-1\rangle\big)+$\\&&$+\frac{1}{3
{\sqrt{5}}}\big(|-1,1,0,0\rangle+
|0,0,-1,1\rangle+|0,0,1,-1\rangle+
|1,-1,0,0\rangle\big)+$\\&&$+\frac{2}{3
{\sqrt{5}}}|0,0,0,0\rangle+
\frac{1}{{\sqrt{5}}}\big(|-1,-1,1,1\rangle+|1,1,-1,-1\rangle\big);$\\

4&2&$-\frac{1}{2
{\sqrt{3}}}\big(|-1,0,1,0\rangle+|-1,1,-1,1\rangle+
|0,-1,0,1\rangle+|0,1,0,-1\rangle+$\\&&$+
|1,-1,1,-1\rangle+|1,0,-1,0\rangle\big)+\frac{1}{2
{\sqrt{3}}}\big(|-1,0,0,1\rangle+|-1,1,1,-1\rangle+$\\&&$+
|0,-1,1,0\rangle+|0,1,-1,0\rangle+
|1,-1,-1,1\rangle+|1,0,0,-1\rangle\big);$\\
4&3&$\big(\frac{1}{{\sqrt{3}}}\big(-|0,0\rangle+|-1,1\rangle+|1,-1\rangle\big)\big)^2
;$\\\hline 5&1&$-{\sqrt{\frac{2}{15}}}|-1,-1,0,1,1\rangle+
-\frac{1}{{\sqrt{30}}}\big(|-1,0,1,0,0\rangle+|0,-1,1,0,0\rangle+$\\&&$+
|0,0,-1,0,1\rangle+|0,0,-1,1,0\rangle+
|0,1,1,-1,-1\rangle+$\\&&$+|1,0,1,-1,-1\rangle+
|1,1,-1,-1,0\rangle+|1,1,-1,0,-1\rangle\big)+$\\&&$+ -\frac{1}{2
{\sqrt{30}}}\big(|-1,0,1,-1,1\rangle+|-1,0,1,1,-1\rangle+
|-1,1,-1,0,1\rangle+$\\&&$+|-1,1,-1,1,0\rangle+
|0,-1,1,-1,1\rangle+|0,-1,1,1,-1\rangle+$\\&&$+
|0,1,0,-1,0\rangle+|0,1,0,0,-1\rangle+
|1,-1,-1,0,1\rangle+$\\&&$+|1,-1,-1,1,0\rangle+
|1,0,0,-1,0\rangle+|1,0,0,0,-1\rangle\big)+$\\&&$+ \frac{1}{2
{\sqrt{30}}}\big(|-1,0,0,0,1\rangle+|-1,0,0,1,0\rangle+
|-1,1,1,-1,0\rangle+$\\&&$+|-1,1,1,0,-1\rangle+
|0,-1,0,0,1\rangle+|0,-1,0,1,0\rangle+$\\&&$+
|0,1,-1,-1,1\rangle+|0,1,-1,1,-1\rangle+
|1,-1,1,-1,0\rangle+$\\&&$+|1,-1,1,0,-1\rangle+
|1,0,-1,-1,1\rangle+|1,0,-1,1,-1\rangle\big)+$\\&&$+
\frac{1}{{\sqrt{30}}}\big(|-1,-1,1,0,1\rangle+|-1,-1,1,1,0\rangle+
|-1,0,-1,1,1\rangle+$\\&&$+|0,-1,-1,1,1\rangle+
|0,0,1,-1,0\rangle+|0,0,1,0,-1\rangle+$\\&&$+
|0,1,-1,0,0\rangle+|1,0,-1,0,0\rangle\big)+{\sqrt{\frac{2}{15}}}|1,1,0,-1,-1\rangle;$\\
\hline\hline
\end{tabular}
\caption{First singlet states of $N$ spin one particles.
\label{2005-singlet-t1}}
\end{table}

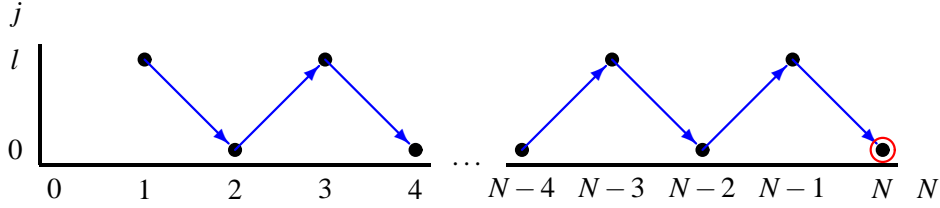
\begin{figure}
\begin{center}
\unitlength 0.40mm
\allinethickness{1pt} 
\begin{picture}(305.33,150.00)
\put(15.00,2.00){\makebox(0,0)[cc]{$0$}}
\put(45.00,2.00){\makebox(0,0)[cc]{$1$}}
\put(75.00,2.00){\makebox(0,0)[cc]{$2$}}
\put(105.00,2.00){\makebox(0,0)[cc]{$3$}}
\put(135.00,2.00){\makebox(0,0)[cc]{$4$}}
\put(2.00,15.00){\makebox(0,0)[cc]{${0}$}}
\put(2.00,45.00){\makebox(0,0)[cc]{${l}$}}
\put(2.00,60.00){\makebox(0,0)[cc]{$j$}}
\put(10.00,10.00){\line(0,1){40.00}}
\put(10.00,10.00){\line(1,0){130.00}}
\put(45.00,45.00){\color{blue} \circle*{4.00}}
\put(75.00,15.00){\color{blue} \circle*{4.00}}
\put(105.00,45.00){\color{blue} \circle*{4.00}}
\put(135.00,15.00){\color{blue} \circle*{4.00}}
\put(45.00,45.00){\color{blue} \vector(1,-1){28.00}}
\put(75.00,15.00){\color{blue} \vector(1,1){28.00}}
\put(105.00,45.00){\color{blue} \vector(1,-1){28.00}}
\put(170.33,2.00){\makebox(0,0)[cc]{$N-4$}}
\put(200.33,2.00){\makebox(0,0)[cc]{$N-3$}}
\put(230.33,2.00){\makebox(0,0)[cc]{$N-2$}}
\put(260.33,2.00){\makebox(0,0)[cc]{$N-1$}}
\put(290.33,2.00){\makebox(0,0)[cc]{$N$}}
\put(305.33,2.00){\makebox(0,0)[cc]{$N$}}
\put(165.33,10.00){\line(1,0){130.00}}
\put(170.33,15.00){\color{blue} \circle*{4.00}}
\put(200.33,45.00){\color{blue} \circle*{4.00}}
\put(230.33,15.00){\color{blue} \circle*{4.00}}
\put(260.33,45.00){\color{blue} \circle*{4.00}}
\put(290.33,15.00){\color{blue} \circle*{4.00}}
\put(170.33,15.00){\color{blue} \vector(1,1){28.00}}
\put(200.33,45.00){\color{blue} \vector(1,-1){28.00}}
\put(230.33,15.00){\color{blue} \vector(1,1){28.00}}
\put(260.33,45.00){\color{blue} \vector(1,-1){28.00}}
\put(290.33,15.00){\color{red} \circle{8.00}}
\put(152.08,10.00){\makebox(0,0)[cc]{$\cdots$}}
\end{picture}
\end{center}
\caption{ Construction of the ``zigzag'' singlet state of $N$
particles which effectively is a product state of $\frac{N}{2}$
spin $l$ particle states. \label{2005-singlet-f1-zigzag}}
\end{figure}

There always exist trivial ``zigzag'' singlet states which are
the product of $r$ two-particle singlet states stemming from the
rising and lowering of consecutive states. The situation is
depicted in Fig.~\ref{2005-singlet-f1-zigzag}. For $j=1$ and
$N=3r$ there exist ``zigzag'' singlet states, which are the
product of $r$ three-particle singlet states. For singlet states
with $N=2r+3t$ ($r,t$ integer) there exist singlet states being
the product of $r$
two-particle singlet states and $t$ three-particle singlet states.

\section{Symmetries}

In what follows we shall discuss the symmetry behavior of singlet
states.
In our approach the singlet states are orthogonal to each other.
This can be demonstrated by considering the formula~\cite{Hagedorn}
\begin{equation}
\begin{array}{rcl}
\langle(j_1'j_2')jm|(j_1j_2)jm\rangle &=&\sum_{m_1'+m_2'=m,
m_1+m_2=m}\langle(j_1'j_2')jm|j_1'm_1'j_2'm_2'\rangle\times \\
&&\langle j_1'm_1'j_2'm_2'|j_1m_1j_2m_2\rangle\langle
j_1m_1j_2m_2|(j_1j_2)jm\rangle\\
&=&\delta_{j_1j_1'}\delta_{j_2j_2'}\delta_{m_1m_1'}\delta_{m_2m_2'},
\end{array}
\end{equation}
where $|(j_1j_2)jm\rangle$ stands for a state of total angular momentum $j$ and
magnetic quantum number $m$, composed of two parts having angular momentum $j_1$ and $j_2$, respectively.
States stemming from different
$j_1$ values are orthogonal to each other. Hence, also the singlet
states derived from them are orthogonal. By iteration it follows
that even singlet states stemming from the same $j_1$ are
orthogonal.
The method allows us to construct the full basis for each singlet space
which has the appropriate dimension.

\subsection{Sign changes of magnetic quantum numbers}


For the Clebsch-Gordan coefficients the following formula holds
\begin{equation}
\langle j_1,-m_1,j_2,-m_2|j,-m\rangle
=(-1)^{j_1+j_2-j}\langle j_1m_1j_2m_2|jm\rangle .
\end{equation}

\subsubsection{Spin one-half}%
In what follows, the symmetries of singlet spin one-half particle states are investigated.
For a coupling $j$ to $j+\frac{1}{2}$, the Clebsch-Gordan
coefficients satisfy
\begin{equation}
\begin{array}{rcl}
\langle
j,-m-\frac{1}{2},\frac{1}{2},\frac{1}{2}|j+\frac{1}{2},-m\rangle &=&(-1)^{0}\langle j,m+\frac{1}{2},\frac{1}{2},-\frac{1}{2}|j+\frac{1}{2},m\rangle\nonumber \\
\langle
j,m+\frac{1}{2},\frac{1}{2},-\frac{1}{2}|j+\frac{1}{2},m\rangle &=&(-1)^{0}\langle j,-m-\frac{1}{2},\frac{1}{2},\frac{1}{2}|j+\frac{1}{2},-m\rangle.
\end{array}
\end{equation}
If  all the magnetic quantum numbers reverse their signs, the
Clebsch-Gordan coefficients stay the same. Coupling
$j+\frac{1}{2}$ to $j$ results in
\begin{equation}
\begin{array}{rcl}
\langle
j+\frac{1}{2},m,\frac{1}{2},\frac{1}{2}|j,m+\frac{1}{2}\rangle &=&(-1)^{1}\langle j+\frac{1}{2},-m,\frac{1}{2},-\frac{1}{2}|j,-m-\frac{1}{2}\rangle\nonumber\\
\langle
j+\frac{1}{2},-m,\frac{1}{2},-\frac{1}{2}|j,-m-\frac{1}{2}\rangle  &=&(-1)^{1}\langle j+\frac{1}{2},m,\frac{1}{2},\frac{1}{2}|j,m+\frac{1}{2}\rangle.
\end{array}
\end{equation}
In this case, all the Clebsch-Gordan coefficients change their signs.

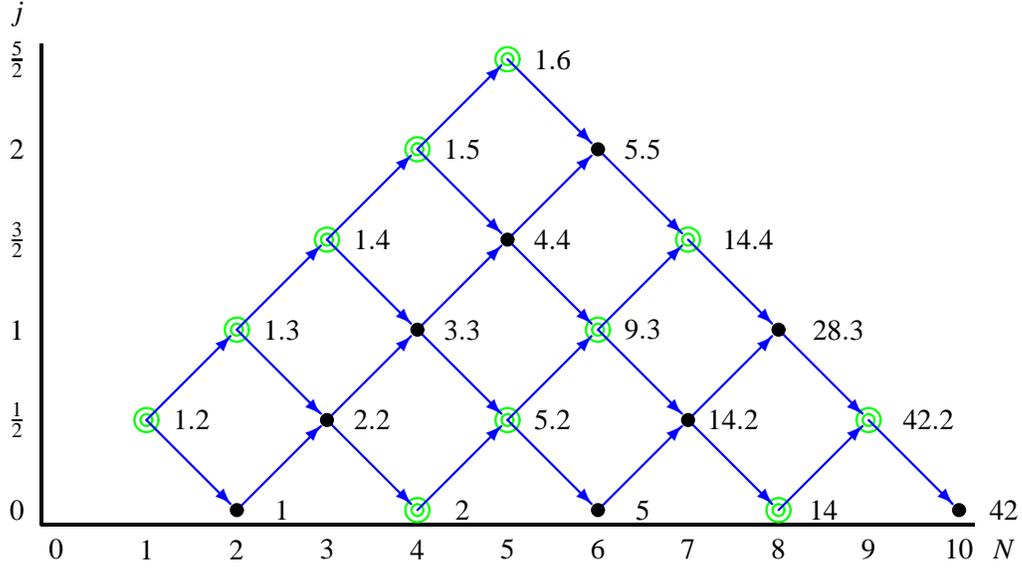
\begin{figure}
\begin{center}
\unitlength 0.40mm
\allinethickness{1pt} 
\begin{picture}(350.00,200.00)
\put(15.00,2.00){\makebox(0,0)[cc]{$0$}}
\put(45.00,2.00){\makebox(0,0)[cc]{$1$}}
\put(75.00,2.00){\makebox(0,0)[cc]{$2$}}
\put(105.0,2.00){\makebox(0,0)[cc]{$3$}}
\put(135.00,2.00){\makebox(0,0)[cc]{$4$}}
\put(165.00,2.00){\makebox(0,0)[cc]{$5$}}
\put(195.00,2.00){\makebox(0,0)[cc]{$6$}}
\put(225.0,2.00){\makebox(0,0)[cc]{$7$}}
\put(255.00,2.00){\makebox(0,0)[cc]{$8$}}
\put(285.0,2.00){\makebox(0,0)[cc]{$9$}}
\put(315.00,2.00){\makebox(0,0)[cc]{$10$}}
\put(330.00,2.00){\makebox(0,0)[cc]{$N$}}
\put(2.00,15.00){\makebox(0,0)[cc]{$0$}}
\put(2.00,45.00){\makebox(0,0)[cc]{${1\over 2}$}}
\put(2.00,75.00){\makebox(0,0)[cc]{$1$}}
\put(2.00,105.00){\makebox(0,0)[cc]{${3\over 2}$}}
\put(2.00,135.00){\makebox(0,0)[cc]{$2$}}
\put(2.00,165.00){\makebox(0,0)[cc]{${5\over 2}$}}
\put(2.00,180.00){\makebox(0,0)[cc]{$j$}}
\put(10.00,10.00){\line(0,1){160.00}}
\put(10.00,10.00){\line(1,0){310.00}}
%
\put(135.00,15.00){\color{green}\circle{4.00}}
\put(135.00,15.00){\color{green} \circle{8.00}}
\put(255.00,15.00){\color{green}\circle{4.00}}
\put(255.00,15.00){\color{green}\circle{8.00}}
\put(45.00,45.00){\color{green}\circle{4.00}}
\put(45.00,45.00){\color{green}\circle{8.00}}
\put(165.00,45.00){\color{green}\circle{4.00}}
\put(165.00,45.00){\color{green}\circle{8.00}}
\put(285.00,45.00){\color{green}\circle{4.00}}
\put(285.00,45.00){\color{green}\circle{8.00}}
\put(75.00,75.00){\color{green}\circle{4.00}}
 \put(75.00,75.00){\color{green}\circle{8.00}}
\put(195.00,75.00){\color{green}\circle{4.00}}
\put(195.00,75.00){\color{green}\circle{8.00}}
\put(105.00,105.00){\color{green}\circle{4.00}}
\put(105.00,105.00){\color{green}\circle{8.00}}
\put(225.00,105.00){\color{green}\circle{4.00}}
\put(225.00,105.00){\color{green}\circle{8.00}}
\put(135.00,135.00){\color{green}\circle{4.00}}
\put(135.00,135.00){\color{green}\circle{8.00}}
\put(165.00,165.00){\color{green}\circle{4.00}}
\put(165.00,165.00){\color{green}\circle{8.00}}
\put(75.00,15.00){\color{blue} \vector(1,1){28.00}}
\put(135.00,15.00){\color{blue} \vector(1,1){28.00}}
\put(195.00,15.00){\color{blue} \vector(1,1){28.00}}
\put(255.00,15.00){\color{blue} \vector(1,1){28.00}}
\put(45.00,45.00){\color{blue} \vector(1,-1){28.00}}
\put(45.00,45.00){\color{blue} \vector(1,1){28.00}}
\put(105.00,45.00){\color{blue} \vector(1,-1){28.00}}
\put(105.00,45.00){\color{blue} \vector(1,1){28.00}}
\put(165.00,45.00){\color{blue} \vector(1,-1){28.00}}
\put(165.00,45.00){\color{blue} \vector(1,1){28.00}}
\put(225.00,45.00){\color{blue} \vector(1,-1){28.00}}
\put(225.00,45.00){\color{blue} \vector(1,1){28.00}}
\put(285.00,45.00){\color{blue} \vector(1,-1){28.00}}
\put(75.00,75.00){\color{blue} \vector(1,-1){28.00}}
\put(75.00,75.00){\color{blue} \vector(1,1){28.00}}
\put(135.00,75.00){\color{blue} \vector(1,-1){28.00}}
\put(135.00,75.00){\color{blue} \vector(1,1){28.00}}
\put(195.00,75.00){\color{blue} \vector(1,-1){28.00}}
\put(195.00,75.00){\color{blue} \vector(1,1){28.00}}
\put(255.00,75.00){\color{blue} \vector(1,-1){28.00}}
\put(105.00,105.00){\color{blue} \vector(1,-1){28.00}}
\put(105.00,105.00){\color{blue} \vector(1,1){28.00}}
\put(165.00,105.00){\color{blue} \vector(1,-1){28.00}}
\put(165.00,105.00){\color{blue} \vector(1,1){28.00}}
\put(225.00,105.00){\color{blue} \vector(1,-1){28.00}}
\put(135.00,135.00){\color{blue} \vector(1,-1){28.00}}
\put(135.00,135.00){\color{blue} \vector(1,1){28.00}}
\put(195.00,135.00){\color{blue} \vector(1,-1){28.00}}
\put(165.00,165.00){\color{blue} \vector(1,-1){28.00}}
\put(75.00,15.00){\color{magenta}\circle*{4.00}}
\put(195.00,15.00){\color{magenta}\circle*{4.00}}
\put(315.00,15.00){\color{magenta}\circle*{4.00}}
\put(105.00,45.00){\color{magenta}\circle*{4.00}}
\put(225.00,45.00){\color{magenta}\circle*{4.00}}
\put(135.00,75.00){\color{magenta}\circle*{4.00}}
\put(255.00,75.00){\color{magenta}\circle*{4.00}}
\put(165.00,105.00){\color{magenta}\circle*{4.00}}
\put(195.00,135.00){\color{magenta}\circle*{4.00}}
\put(85,15){\makebox(10,0){\it $1$}}
 \put(145,15){\makebox(10,0){\bf $2$}}
 \put(205,15){\makebox(10,0){\it $5$}}
\put(265,15){\makebox(10,0){\bf $14$}}
 \put(325,15){\makebox(10,0){\it $42$}}
 \put(55,45){\makebox(10,0){\bf $1.2$}}
\put(115,45){\makebox(10,0){\it $2.2$}}
 \put(175,45){\makebox(10,0){\bf $5.2$}}
 \put(235,45){\makebox(10,0){\it $14.2$}}
\put(300,45){\makebox(10,0){\bf $42.2$}}
 \put(85,75){\makebox(10,0){\bf $1.3$}}
 \put(145,75){\makebox(10,0){\it $3.3$}}
\put(205,75){\makebox(10,0){\bf $9.3$}}
 \put(270,75){\makebox(10,0){\it $28.3$}}
 \put(115,105){\makebox(10,0){\bf $1.4$}}
\put(175,105){\makebox(10,0){\it $4.4$}}
 \put(240,105){\makebox(10,0){\bf $14.4$}}
 \put(145,135){\makebox(10,0){\bf $1.5$}}
\put(205,135){\makebox(10,0){\it $5.5$}}
\put(175,165){\makebox(10,0){\bf $1.6$}}
\end{picture}
\end{center}
\caption{Symmetry behavior of spin one-half particles.
\label{2005-singlet-f1-ta-tako} Even and odd subspaces are denoted by
concentric and filled circles, respectively.
The numbers denote the dimensions of the subspaces. The first
number stands for the number of states $|h,j\rangle$, and the
second number stands for the $2j+1$ projections. Arrows represent the direction
of the coupling.}
\end{figure}

We conclude that the symmetry behavior remains the same if one passes
from the angular momentum subspace $|N,J\rangle$ to the angular
momentum subspace $|N+1,J+{1\over 2}\rangle$. By passing from the
subspace $|N,J\rangle$ to the subspace $|N+1,J-{1\over2}\rangle$
the symmetry behaviour changes from even to odd and from odd to
even, respectively.
A graphical representation of this property is depicted in Fig.~\ref{2005-singlet-f1-ta-tako}.
In particular, the singlet states where $N$ is $k\cdot 2\cdot 2$ (k is an integer) are even, and the ones where $N$ is $ 2\cdot (2k+1)$ are odd.

\subsubsection{Spin one}%

Let us now
consider the $j=1$ case first. For the coupling of $j$ to $j+1$, the symmetry described above
implies
\begin{equation}
\begin{array}{rcl}
\langle j,-m-1,1,1|j+1,-m\rangle
&= &(-1)^{0}\langle j,m+1,1,-1|j+1,m\rangle ,\\
\langle j,-m,1,0|j+1,-m\rangle &= &(-1)^{0}\langle
j,m,1,0|j+1,m\rangle ;
\end{array}
\end{equation}
i.e., the Clebsch-Gordan
coefficients are the same.
For the coupling of $j$ to $j$,
\begin{equation}
\begin{array}{rcl}
\langle j,-m-1,1,1|j,-m\rangle
&= &(-1)^{1}\langle j,m+1,1,-1|j,m\rangle ,\\
\langle j,-m,1,0|j,-m\rangle & = &(-1)^{1}\langle
j,m,1,0|j,m\rangle ;
\end{array}
\end{equation}
i.e., all Clebsch-Gordan coefficients change  sign.
Similarly for the coupling of $j+1$ to $j$,
\begin{equation}
\begin{array}{rcl}
\langle j+1,m,1,1|j,m+1\rangle
&= &(-1)^{2}\langle j+1,-m,1,-1|j,-m-1\rangle ,\\
\langle j+1,m,1,0|j,m\rangle &= &(-1)^{2}\langle
j+1,-m,1,0|j,-m\rangle ;
\end{array}
\end{equation}
i.e., they all stay the same.

\begin{figure}
\begin{center}
\unitlength 0.50mm
\allinethickness{1pt} 
\begin{picture}(290.00,150.00)
\put(15.00,2.00){\makebox(0,0)[cc]{$0$}}
\put(45.00,2.00){\makebox(0,0)[cc]{$1$}}
\put(75.00,2.00){\makebox(0,0)[cc]{$2$}}
\put(105.0,2.00){\makebox(0,0)[cc]{$3$}}
\put(135.00,2.00){\makebox(0,0)[cc]{$4$}}
\put(165.00,2.00){\makebox(0,0)[cc]{$5$}}
\put(195.00,2.00){\makebox(0,0)[cc]{$6$}}
\put(225.00,2.00){\makebox(0,0)[cc]{$7$}}
\put(255.0,2.00){\makebox(0,0)[cc]{$8$}}
\put(285.00,2.00){\makebox(0,0)[cc]{$9$}}
\put(315.00,2.00){\makebox(0,0)[cc]{$10$}}
\put(330.00,2.00){\makebox(0,0)[cc]{$N$}}
\put(2.00,15.00){\makebox(0,0)[cc]{$0$}}
\put(2.00,45.00){\makebox(0,0)[cc]{$1$}}
\put(2.00,75.00){\makebox(0,0)[cc]{$2$}}
\put(2.00,105.00){\makebox(0,0)[cc]{$3$}}
\put(2.00,135.00){\makebox(0,0)[cc]{$4$}}
\put(2.00,165.00){\makebox(0,0)[cc]{$5$}}
\put(2.00,180.00){\makebox(0,0)[cc]{$j$}}
\put(10.00,10.00){\line(0,1){160.00}}
\put(10.00,10.00){\line(1,0){310.00}}

\put(75.00,15.00){\color{green}\circle{4.00}}
\put(75.00,15.00){\color{green}\circle{8.00}}
 \put(135.00,15.00){\color{green}\circle{4.00}}
\put(135.00,15.00){\color{green}\circle{8.00}}
\put(195.00,15.00){\color{green}\circle{4.00}}
 \put(195.00,15.00){\color{green}\circle{8.00}}
  \put(255.00,15.00){\color{green}\circle{4.00}}
\put(255.00,15.00){\color{green}\circle{8.00}}
\put(315.00,15.00){\color{green}\circle{4.00}}
 \put(315.00,15.00){\color{green}\circle{8.00}}
\put(45.00,45.00){\color{green}\circle{4.00}}
 \put(45.00,45.00){\color{green}\circle{8.00}}
\put(105.00,45.00){\color{green}\circle{4.00}}
\put(105.00,45.00){\color{green}\circle{8.00}}
\put(165.00,45.00){\color{green}\circle{4.00}}
 \put(165.00,45.00){\color{green}\circle{8.00}}
\put(225.00,45.00){\color{green}\circle{8.00}}
 \put(225.00,45.00){\color{green}\circle{4.00}}
 \put(285.00,45.00){\color{green}\circle{4.00}}
  \put(285.00,45.00){\color{green}\circle{8.00}}
\put(75.00,75.00){\color{green}\circle{4.00}}
\put(75.00,75.00){\color{green}\circle{8.00}}
 \put(135.00,75.00){\color{green}\circle{4.00}}
\put(135.00,75.00){\color{green}\circle{8.00}}
\put(195.00,75.00){\color{green}\circle{4.00}}
 \put(195.00,75.00){\color{green}\circle{8.00}}
\put(255.00,75.00){\color{green}\circle{4.00}}
 \put(255.00,75.00){\color{green}\circle{8.00}}
\put(105.00,105.00){\color{green}\circle{4.00}}
\put(105.00,105.00){\color{green}\circle{8.00}}
\put(165.00,105.00){\color{green}\circle{4.00}}
 \put(165.00,105.00){\color{green}\circle{8.00}}
\put(225.00,105.00){\color{green}\circle{8.00}}
 \put(225.00,105.00){\color{green}\circle{4.00}}
  \put(135.00,135.00){\color{green}\circle{4.00}}
\put(135.00,135.00){\color{green}\circle{8.00}}
\put(195.00,135.00){\color{green}\circle{4.00}}
 \put(195.00,135.00){\color{green}\circle{8.00}}
\put(165.00,165.00){\color{green}\circle{4.00}}
 \put(165.00,165.00){\color{green}\circle{8.00}}
\put(75,15){\color{blue}\vector(1,1){28}}
\put(105,15){\color{blue}\vector(1,1){28}}\put(135,15){\color{blue}\vector(1,1){28}}
\put(165,15){\color{blue}\vector(1,1){28}}\put(195,15){\color{blue}\vector(1,1){28}}
\put(225,15){\color{blue}\vector(1,1){28}}\put(255,15){\color{blue}\vector(1,1){28}}
 \put(45,45){\color{blue}\vector(1,1){28}}
\put(45,45){\color{blue}\vector(1,0){28}} \put(45,45){\color{blue}\vector(1,-1){28}}
\put(75,45){\color{blue}\vector(1,1){28}} \put(75,45){\color{blue}\vector(1,0){28}}
\put(75,45){\color{blue}\vector(1,-1){28}} \put(105,45){\color{blue}\vector(1,1){28}}
\put(105,45){\color{blue}\vector(1,0){28}}
\put(105,45){\color{blue}\vector(1,-1){28}}\put(135,45){\color{blue}\vector(1,1){28}}
\put(135,45){\color{blue}\vector(1,0){28}}
\put(135,45){\color{blue}\vector(1,-1){28}}\put(165,45){\color{blue}\vector(1,1){28}}
\put(165,45){\color{blue}\vector(1,0){28}} \put(165,45){\color{blue}\vector(1,-1){28}}
\put(195,45){\color{blue}\vector(1,1){28}} \put(195,45){\color{blue}\vector(1,0){28}}
\put(195,45){\color{blue}\vector(1,-1){28}} \put(225,45){\color{blue}\vector(1,1){28}}
\put(225,45){\color{blue}\vector(1,0){28}} \put(225,45){\color{blue}\vector(1,-1){28}}
\put(255,45){\color{blue}\vector(1,0){28}} \put(255,45){\color{blue}\vector(1,-1){28}}
\put(285,45){\color{blue}\vector(1,-1){28}}
 \put(75,75){\color{blue}\vector(1,1){28}}
\put(75,75){\color{blue}\vector(1,0){28}} \put(75,75){\color{blue}\vector(1,-1){28}}
\put(105,75){\color{blue}\vector(1,1){28}} \put(105,75){\color{blue}\vector(1,0){28}}
\put(105,75){\color{blue}\vector(1,-1){28}}\put(135,75){\color{blue}\vector(1,1){28}}
\put(135,75){\color{blue}\vector(1,0){28}}
\put(135,75){\color{blue}\vector(1,-1){28}}\put(165,75){\color{blue}\vector(1,1){28}}
\put(165,75){\color{blue}\vector(1,0){28}} \put(165,75){\color{blue}\vector(1,-1){28}}
\put(195,75){\color{blue}\vector(1,1){28}} \put(195,75){\color{blue}\vector(1,0){28}}
\put(195,75){\color{blue}\vector(1,-1){28}} \put(225,75){\color{blue}\vector(1,0){28}}
\put(225,75){\color{blue}\vector(1,-1){28}} \put(225,75){\color{blue}\vector(1,-1){28}}
\put(255,75){\color{blue}\vector(1,-1){28}} \put(105,105){\color{blue}\vector(1,1){28}}
\put(105,105){\color{blue}\vector(1,0){28}}
\put(105,105){\color{blue}\vector(1,-1){28}}\put(135,105){\color{blue}\vector(1,1){28}}
\put(135,105){\color{blue}\vector(1,0){28}}
\put(135,105){\color{blue}\vector(1,-1){28}}\put(165,105){\color{blue}\vector(1,1){28}}
\put(165,105){\color{blue}\vector(1,0){28}} \put(165,105){\color{blue}\vector(1,-1){28}}
\put(195,105){\color{blue}\vector(1,0){28}} \put(195,105){\color{blue}\vector(1,-1){28}}
\put(195,105){\color{blue}\vector(1,-1){28}} \put(225,105){\color{blue}\vector(1,-1){28}}
\put(135,135){\color{blue}\vector(1,1){28}} \put(135,135){\color{blue}\vector(1,0){28}}
\put(135,135){\color{blue}\vector(1,-1){28}} \put(165,135){\color{blue}\vector(1,0){28}}
\put(165,135){\color{blue}\vector(1,-1){28}}
\put(195,135){\color{blue}\vector(1,-1){28}}\put(165,165){\color{blue}\vector(1,-1){28}}

\put(105.00,15.00){\color{magenta} \circle*{4.00}}
 \put(165.00,15.00){\color{magenta}\circle*{4.00}}
 \put(225.00,15.00){\color{magenta}\circle*{4.00}}
 \put(285.00,15.00){\color{magenta}\circle*{4.00}}
 \put(75.00,45.00){\color{magenta}\circle*{4.00}}
 \put(135.00,45.00){\color{magenta}\circle*{4.00}}
 \put(195.00,45.00){\color{magenta}\circle*{4.00}}
\put(255.00,45.00){\color{magenta}\circle*{4.00}}
\put(105.00,75.00){\color{magenta}\circle*{4.00}}
 \put(165.00,75.00){\color{magenta}\circle*{4.00}}
 \put(225.00,75.00){\color{magenta}\circle*{4.00}}
 \put(135.00,105.00){\color{magenta}\circle*{4.00}}
  \put(195.00,105.00){\color{magenta}\circle*{4.00}}
 \put(165.00,135.00){\color{magenta}\circle*{4.00}}

\put(81, 16)   {\makebox(10,5){\bf $1$}}
\put(111,16)  {\makebox(10,5){\it $1$}}
\put(141,16)  {\makebox(10,5){\bf $3$}}
\put(171,16)  {\makebox(10,5){\it $6$}}
\put(203,16)  {\makebox(10,5){\bf $15$}}
\put(233,16)  {\makebox(10,5){\it $36$}}
\put(263,16)  {\makebox(10,5){\bf $91$}}
\put(293,16)  {\makebox(10,5){\it $232$}}
\put(324,16)  {\makebox(10,5){\bf $603$}}
\put(53, 46)   {\makebox(10,7){\bf $1.3$}}
\put(83, 46)   {\makebox(10,7){\it $1.3$}}
\put(113,46)  {\makebox(10,7){\bf $3.3$}}
\put(143,46)  {\makebox(10,7){\it $6.3$}}
\put(173,46)  {\makebox(10,7){\bf $15.3$}}
\put(203,46)  {\makebox(10,7){\it $36.3$}}
\put(233,46)  {\makebox(10,7){\bf $91.3$}}
\put(263,46)  {\makebox(10,7){\it $232.3$}}
\put(296,46)  {\makebox(10,7){\bf $603.3$}}
\put(83 ,76)   {\makebox(10,7){\bf $1.5$}}
\put(113,76)  {\makebox(10,7){\it $2.5$}}
\put(143,76)  {\makebox(10,7){\bf $6.5$}}
\put(173,76)  {\makebox(10,7){\it $15.5$}}
\put(203,76)  {\makebox(10,7){\bf $40.5$}}
\put(233,76)  {\makebox(10,7){\it $105.5$}}
\put(265,76)  {\makebox(10,7){\bf $280.5$}}
\put(113,106){\makebox(10,7){\bf $1.7$}}
\put(143,106){\makebox(10,7){\it $3.7$}}
\put(173,106){\makebox(10,7){\bf $10.7$}}
\put(203,106){\makebox(10,7){\it $29.7$}}
\put(233,106){\makebox(10,7){\bf $84.7$}}
\put(143,136){\makebox(10,7){\bf $1.9$}}
\put(173,136){\makebox(10,7){\it $4.9$}}
\put(203,136){\makebox(10,7){\bf $15.9$}}
\put(173,166){\makebox(10,7){\bf $1.11$}}


\end{picture}
\end{center}
\caption{Symmetries of spin one particle states.
\label{2005-singlet-f1-ta-takohalf} Even subspaces are denoted by
concentric circles, odd subspaces are denoted by filled circles.
The numbers denote the dimensions of the subspaces. The first
number stands for the number of states $|h,j\rangle$ and the
second stands for the $2j+1$ projections. Arrows represent the way
of coupling. }
\end{figure}
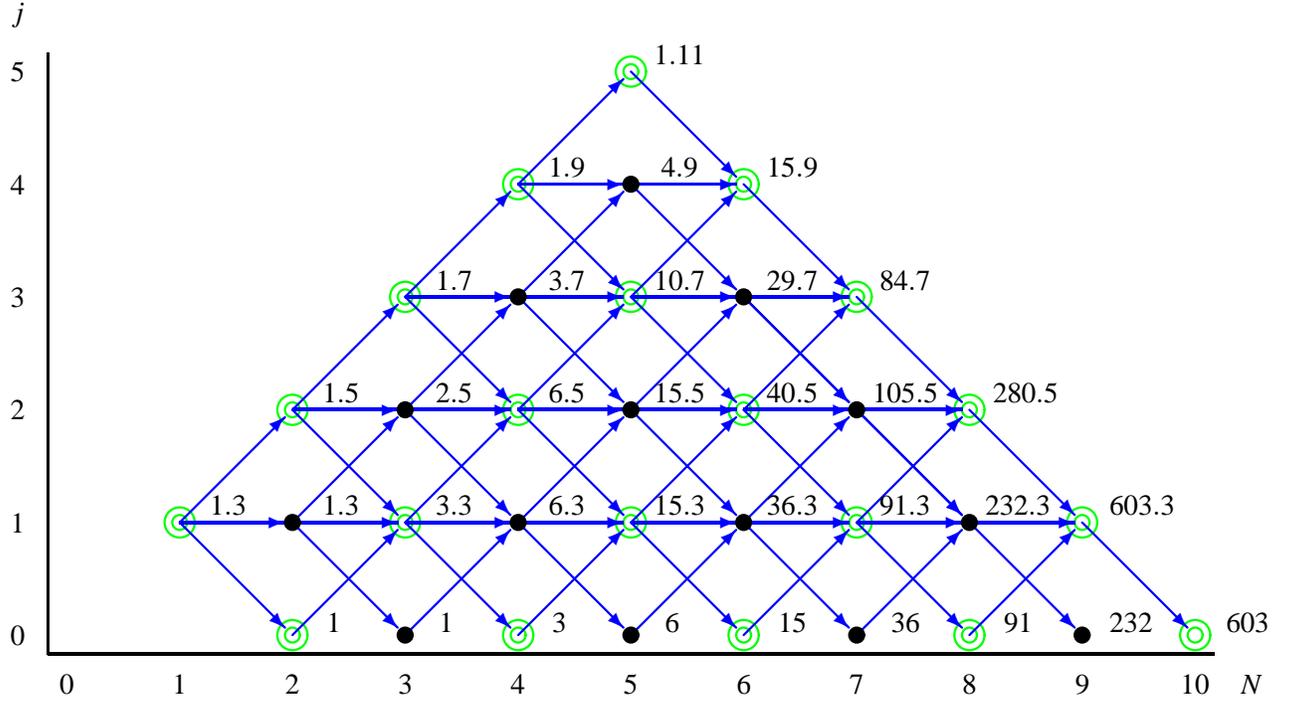

Using these symmetries, we conclude that the symmetry behaviour
remains the same if one passes from the angular momentum subspace
$|N,j\rangle$ to the angular momentum subspace $|N+1,j+1\rangle$.
The symmetry behaviour does not change for coupling $|N,j+1\rangle$
to $|N+1,j\rangle$.
Coupling $|N,j\rangle$ to $|N+1,j\rangle$ changes the symmetry behaviour from even to
odd and from odd to even.
The situation is depicted in
Fig.~\ref{2005-singlet-f1-ta-takohalf}.
$N$-particle singlet states with $N$ even are even,
whereas $N$-particle singlet states with $N$ odd are odd.

\subsection{Symmetric group}

Let us consider the permutations of the $N$ magnetic quantum numbers
in every product state of  $N$ particles.
More explicitly, since every
permutation of $N$ particles can be written as the product of
$N-1$ transpositions, we shall study the effects of $N-1$ transpositions.
We analyze $N-1$ transpositions of the
form $(j,j+1)$, the transposition of $j$ and $j+1$ which generate
the whole symmetric group,
and in particular all the $N\cdot (N-1)/2$
transpositions, since $(j,k+1)=(k,k+1)(j,k)(k,k+1)$.
Therefore, we consider the class $(2\;1^{N-2})$ of all two particle transpositions. Each irreducible
representation can be labeled by an ordered partition of integers
which corresponds to a specific Young diagram.

As stated in App.~D, Sec.~14 of Ref.~\cite{messiah-62}, the
space spanned by the vectors of total spins $(SM)$ formed by $N$
identical spins ${1\over 2}$ is associated with an irreducible
representation of $S_N$, the representation whose Young diagram
corresponds to the partition $[{1\over 2}N+S,{1\over 2}N-S]$ of
the integer $N$.  It is apparent that the Young diagrams for the
irreducible components of the representation of $S_N$ have at most
two lines. For $N>2$, any state contains at least two individual
spins in the same state.
Suppose the state contains the factor
$u_+^{(i)}u_+^{(j)}$; i.e., $m_i, m_j ={1\over 2}$.
Since
$A={1\over 2}(1-(i,j))$ is the antisymmetrizer and
${1\over 2}(1-(i,j))u_+^{(i)}u_+^{(j)}=0$,
it follows that $A|jm\rangle=0$.

Using the theorem mentioned above, the Young diagrams of the
irreducible spaces of the $N$-particle singlet states correspond to
the partitions $[{1\over 2}N,{1\over 2}N]$. Hence the two-particle
singlet state (sometimes referred to as the {\em ``Bell''} state)
is an antisymmetric one-dimensional space. The four-
and six-particle singlet spaces form a two- and a five-dimensional
irreducible space whose Young diagrams are of the form $[2,2]$ and
$[3,3]$.
Using the formula for the dimension of an irreducible
representation having the partition $[\lambda]$ (e.g., Ref.~\cite{wybourne})
\begin{equation}
f^\lambda = n!  \,
\frac{\Pi_{i<j\leq k}(\lambda_i-\lambda_j+j-i)}{\Pi^k_{i=1}(\lambda_i+k-i)!},\label{dim}
\end{equation}
the dimension can be verified.


\section{Four spin one-half particle correlations}

Singlet states $\vert \Psi_{d,n,i} \rangle$ can be labeled by three numbers $d$, $n$ and $i$,
denoting
the number $d$ of outcomes associated with the dimension of Hilbert space per particle,
the number $n$ of participating particles,
and the state count $i$ in an enumeration of all possible singlet states of $n$ particles of spin $j=(d-1)/2$, respectively.
To begin with the analysis of four-partite correlations, consider four spin one-half
particles in one of the two singlet states enumerated in Table~\ref{2005-singlet-t12} and
computed by following the ``paths'' indicated in Fig.~\ref{2005-singlet-f12-e1}; i.e.,
\begin{eqnarray}
\vert \Psi_{2,4,1} \rangle
&=&
{1\over \sqrt{3}}\Bigl[
\vert ++-- \rangle +
\vert --++ \rangle \nonumber \\
&&\qquad
\qquad
-  {1\over 2}
\bigl(
\vert +- \rangle +
\vert -+ \rangle
\bigr)
\bigl(
\vert +- \rangle +
\vert -+ \rangle
\bigr)
\Bigr],
\label{2005-hp-ep24s2}
\\
\vert \Psi_{2,4,2} \rangle
&=&
\left( \vert \Psi_{2,2,1} \rangle \right)^2
=
{1\over 2}
\bigl(
\vert +- \rangle -
\vert -+ \rangle
\bigr)
\bigl(
\vert +- \rangle -
\vert -+ \rangle
\bigr),
\label{2004-gtq-s1}
\end{eqnarray}
where
$\vert \Psi_{2,2,1} \rangle = \frac{1}{\sqrt{ 2}}
\bigl(
\vert +- \rangle -
\vert -+ \rangle
\bigr)
$
is the two particle singlet ``Bell'' state.

These pure states have an explicit vector space representation as orthogonal vectors.
The two states corresponding to spin ``up'' and ``down'' correspond to
$
\vert +\rangle
\equiv {\hat {\bf e}}_1 =(1,0)
$
and
$
\vert -\rangle \equiv {\hat {\bf e}}_2 =(0,1)
$.
Product states can be represented by the tensor or Kronecker product,
which, for two arbitrary vectors ${\bf a}=(a_1,a_2,\ldots ,a_n)$ and ${\bf b} =(b_1,b_2,\ldots ,b_m)$, can be represented by
\begin{equation}
{\bf a} \otimes {\bf b} = (a_1{\bf b},a_2{\bf b},\ldots ,a_n{\bf b}) = (a_1b_1,a_1b_2,\ldots ,a_nb_m).
\end{equation}
Thus, by summing up all product sates, the two singlet states have a vector representation as
\begin{eqnarray} \hat  \Psi_{2,4,1}
&=&
{1\over \sqrt{3}}\Bigl[
{\hat {\bf e}}_1\otimes {\hat {\bf e}}_1\otimes {\hat {\bf e}}_2\otimes {\hat {\bf e}}_2+
{\hat {\bf e}}_2\otimes {\hat {\bf e}}_2\otimes {\hat {\bf e}}_1\otimes {\hat {\bf e}}_1
 \nonumber \\
&&\qquad
-  {1\over \sqrt{2}}\bigl({\hat {\bf e}}_1\otimes {\hat {\bf e}}_2+{\hat {\bf e}}_2\otimes {\hat {\bf e}}_1 \bigr)
\otimes
 {1\over \sqrt{2}}\bigl({\hat {\bf e}}_1\otimes {\hat {\bf e}}_2+{\hat {\bf e}}_2\otimes {\hat {\bf e}}_1 \bigr)
\Bigr]\nonumber \\
&=&\left( 0,0,0,\frac{1}{{\sqrt{3}}},0,
  -\frac{1}{2\,{\sqrt{3}}},-\frac{1}{2\,{\sqrt{3}}},0,
  0,-\frac{1}{2\,{\sqrt{3}}},-\frac{1}{2\,{\sqrt{3}}},
  0,\frac{1}{{\sqrt{3}}},0,0,0\right).
\label{2005-hp-ep24s2v}
\\
\hat  \Psi_{2,4,2}
 &=&
{1\over \sqrt{2}}\bigl({\hat {\bf e}}_1\otimes {\hat {\bf e}}_2-{\hat {\bf e}}_2\otimes {\hat {\bf e}}_1 \bigr)
\otimes
 {1\over \sqrt{2}}\bigl({\hat {\bf e}}_1\otimes {\hat {\bf e}}_2-{\hat {\bf e}}_2\otimes {\hat {\bf e}}_1 \bigr)
\nonumber \\
&=&\left( 0,0,0,0,0,\frac{1}{2},- \frac{1}{2} ,  0,0,- \frac{1}{2} ,\frac{1}{2},0,0,0,0,  0\right),
\label{2005-hp-ep24s1v}
\end{eqnarray}
Their density operators $\rho_{i}$, $i=1,2$
are just the projectors corresponding to the one-dimensional
linear subspaces spanned by
the vectors representing
$ \hat \Psi_{2,4,2}$
and
$\hat \Psi_{2,4,1}$
in Eqs.~(\ref{2005-hp-ep24s1v},~\ref{2005-hp-ep24s2v}); i.\,e.
they are  the dyadic product
\begin{equation}
\rho_{i} = \left[ {\hat \Psi}_{2,4,i}^T {\hat \Psi}_{2,4,i}\right].
\end{equation}

As has been pointed out above, and as $ \hat \Psi_{2,4,2}\cdot \hat \Psi_{2,4,1}=0$
or equivalently $\rho_{\Psi_{2,4,1}}\cdot \rho_{\Psi_{2,4,2}} = 0$,
the singlet states are orthogonal.
The most general form of a four spin one-half
particle singlet state is
thus given by
\begin{equation}
\vert \Psi_{2,4,s} \rangle
= \lambda_1 \;
\vert \Psi_{2,4,1} \rangle
+
\lambda_2   \;
\vert \Psi_{2,4,2} \rangle
\label{2005-hp-ep24smgf}
\end{equation}
with
$
\vert \lambda_1 \vert^2
+
\vert \lambda_2 \vert^2
=1
$,
which can be parameterized by
$\lambda_1= \sin \tau $,
$\lambda_2=\cos \tau $,
such that
for $\tau =0$,
$\vert \Psi_{2,4,s} \rangle
=
\vert \Psi_{2,4,2} \rangle
$,
and
for $\tau =\pi /2$,
$\vert \Psi_{2,4,s} \rangle
=
\vert \Psi_{2,4,1} \rangle
$.

Singlet states are form invariant with respect to arbitrary unitary
transformations in the single-particle Hilbert spaces and thus
also rotationally invariant in configuration space,
in particular under the rotations
$
\vert + \rangle =
e^{ i{\frac{\varphi}{2}} }
\left(
\cos \frac{\theta}{2} \vert +'  \rangle
-
\sin \frac{\theta}{2} \vert -'   \rangle
\right)
$
and
$
\vert - \rangle =
e^{ -i{\frac{\varphi}{2}} }
\left(
\sin \frac{\theta}{2} \vert +'   \rangle
+
\cos \frac{\theta}{2} \vert -'  \rangle
\right)
$
in the spherical coordinates $\theta , \varphi$ defined below
[e.\,g., Ref.~\cite{krenn1}, Eq.~(2), or Ref.~\cite{ba-89}, Eq.~(7--49)].
However, despite this form invariance under rotations,
the states are non-unique in the sense that knowledge
of a spin state observable for one particle is not sufficient
for the simultaneous (counterfactual) determination of
spin state properties for all other three particles
\cite{epr, svozil-2004-vax}.

\subsection{Operators}

In what follows, the operators corresponding to the spin state observables will be enumerated.
Thereby, spherical coordinates represent angles of spin state measurements.
Suppose  that $i$ denotes the $i$'th particle
with $1\le i\le 4$.
Let $\theta_i$ be the polar angle in the $x$--$z$-plane
from the $z$-axis with $0 \le \theta_i \le \pi$,
and $\varphi_i$  the azimuthal angle in the $x$--$y$-plane
from the $x$-axis with $0 \le \varphi_i < 2 \pi$.

For the sake of simplicity, we shall sometimes consider measurements
in the $x$-$z$-plane, for which $\varphi_1=\varphi_2=\varphi_3=\varphi_4=0$.
Because of the spherical symmetry of the singlet state, this is in every aspect
equivalent to a measurement along angles lying in an arbitrary plane.
In such cases the expectation values (the raw, or uncentered,
product moments \cite{gill-03}) are merely functions of the polar angles
$\theta_1 $,
$\theta_2 $,
$\theta_3 $ and
$\theta_4 $,
so the azimuthal angles will be omitted.
For compact notation,
${\hat \theta}$~and~${\hat \varphi}$ will be used to denote the coordinates
$\theta_1 ,\theta_2 ,\theta_3 ,\theta_4$ and
$\varphi_1 ,\varphi_2 ,\varphi_3 ,\varphi_4$,
respectively.

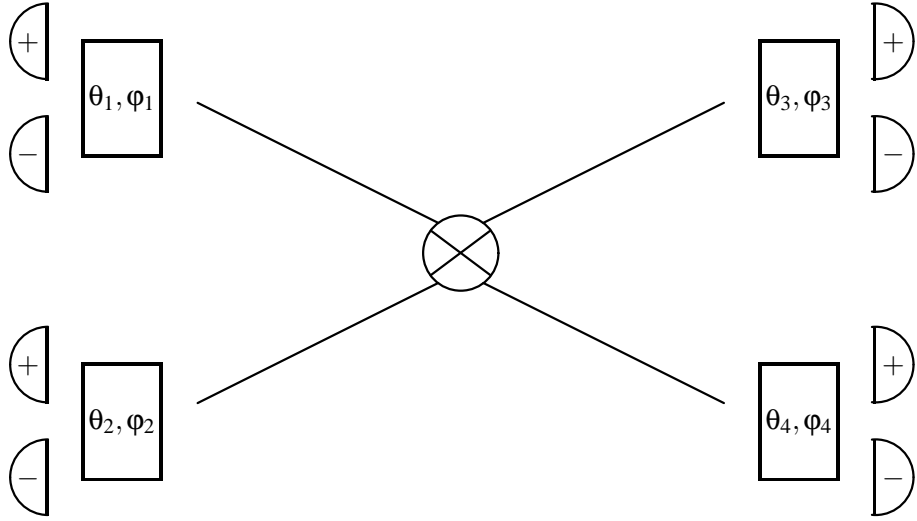
\begin{figure}[htbp]
\begin{center}

\unitlength 1.0mm
\allinethickness{1pt} 
\begin{picture}(120.00,68.00)
\put(60.00,35.00){\color{black} \circle{10.00}}
\put(56.00,32.00){\line(4,3){8.00}}
\put(64.00,32.00){\line(-4,3){8.00}}
\put(57.00,31.00){\line(-2,-1){32.00}}
\put(5.00,5.00){\oval(10.00,10.00)[l]}
\put(5.00,10.00){\line(0,-1){10.00}}
\put(2.50,5.00){\makebox(0,0)[cc]{$-$}}
\put(57.00,39.00){\line(-2,1){32.00}}
\put(5.00,48.00){\oval(10.00,10.00)[l]}
\put(5.00,43.00){\line(0,1){10.00}}
\put(2.50,48.00){\makebox(0,0)[cc]{$-$}}
\put(63.00,31.00){\line(2,-1){32.00}}
\put(63.00,39.00){\line(2,1){32.00}}
\put(5.00,20.00){\oval(10.00,10.00)[l]}
\put(5.00,25.00){\line(0,-1){10.00}}
\put(2.50,20.00){\makebox(0,0)[cc]{$+$}}
\put(5.00,63.00){\oval(10.00,10.00)[l]}
\put(5.00,58.00){\line(0,1){10.00}}
\put(2.50,63.00){\makebox(0,0)[cc]{$+$}}
\put(10.00,5.00){\framebox(10.00,15.00)[cc]{$\theta_2,\varphi_2$}}
\put(10.00,48.00){\framebox(10.00,15.00)[cc]{$\theta_1,\varphi_1$}}
\put(115.00,5.00){\oval(10.00,10.00)[r]}
\put(115.00,10.00){\line(0,-1){10.00}}
\put(117.50,5.00){\makebox(0,0)[cc]{$-$}}
\put(115.00,48.00){\oval(10.00,10.00)[r]}
\put(115.00,43.00){\line(0,1){10.00}}
\put(117.50,48.00){\makebox(0,0)[cc]{$-$}}
\put(115.00,20.00){\oval(10.00,10.00)[r]}
\put(115.00,25.00){\line(0,-1){10.00}}
\put(117.50,20.00){\makebox(0,0)[cc]{$+$}}
\put(115.00,63.00){\oval(10.00,10.00)[r]}
\put(115.00,58.00){\line(0,1){10.00}}
\put(117.50,63.00){\makebox(0,0)[cc]{$+$}}
\put(100.00,5.00){\framebox(10.00,15.00)[cc]{$\theta_4,\varphi_4$}}
\put(100.00,48.00){\framebox(10.00,15.00)[cc]{$\theta_3,\varphi_3$}}
\end{picture}
\end{center}
\caption{Simultaneous spin measurement of
the four-partite singlet state represented in Eq.~(\ref{2004-gtq-s1}).
Boxes indicate spin state analyzers such as Stern-Gerlach apparatus
oriented along the directions $\theta_1,\varphi_1 $,
$\theta_2,\varphi_2 $,
$\theta_3,\varphi_3 $ and
$\theta_4,\varphi_4 $;
their two output ports are occupied with detectors  associated
with the outcomes
``$+$''
and
``$-$'',
respectively.
\label{2005-gtq-f1}}
\end{figure}

The projection operators $F$
corresponding to a four~spin one-half~particle joint measurement
aligned (``$+$'') or antialigned  (``$-$'') along those angles are
\begin{equation}
\begin{array}{lll}
 F_{\pm \pm \pm \pm} ({\hat \theta},{\hat \varphi} ) =
{\frac{1}{2}}\left[{\mathbb I}_2 \pm {\bf \sigma}( \theta_1,\varphi_1 )\right]
\otimes
{\frac{1}{2}}\left[{\mathbb I}_2 \pm {\bf \sigma}( \theta_2,\varphi_2 )\right] \otimes
\nonumber\\
\qquad\qquad\qquad\qquad\qquad
\otimes
{\frac{1}{2}}\left[{\mathbb I}_2 \pm {\bf \sigma}( \theta_3,\varphi_3 )\right]
\otimes
{\frac{1}{2}}\left[{\mathbb I}_2 \pm {\bf \sigma}( \theta_4,\varphi_4 )\right],
\end{array}
\label{2004-gtq-e2}
\end{equation}
with
$
{\bf \sigma}( \theta ,\varphi )=
\left(
\begin{array}{cc} \cos \theta  &e^{-i\varphi} \sin \theta   \\
  e^{i\varphi}\sin \theta  & -\cos \theta
  \end{array}
\right)
$.
For example, $F_{- + - + } ({\hat \theta},{\hat \varphi} )$ stands for the proposition
\begin{quote}
{\em `The spin state of the first particle measured along $\theta_1,\varphi_1$ is ``$-$'',
      the spin state of the second particle measured along $\theta_2,\varphi_2$ is ``$+$'',
      the spin state of the third particle measured along $\theta_3,\varphi_3$ is ``$-$'',
      and the spin state of the fourth particle measured along $\theta_4,\varphi_4$ is ``$+$''~.'
}
\end{quote}
Fig.~\ref{2005-gtq-f1} depicts a measurement configuration
for a simultaneous measurement of spins along
$\theta_1,\varphi_1 $,
$\theta_2,\varphi_2 $,
$\theta_3,\varphi_3 $ and
$\theta_4,\varphi_4 $
of the state $\Psi_{2,4,2}$.

\subsection{Probabilities and expectations}

We now turn to the calculation of quantum predictions.
The joint probability to register the spins of the four particles
in state $\rho_{\Psi_{2,4,s}}$
aligned or antialigned along the directions defined by
($\theta_1$, $\varphi_1 $),
($\theta_2$, $\varphi_2 $),
($\theta_3$, $\varphi_3 $),  and
($\theta_4$, $\varphi_4 $) can be evaluated by a straightforward calculation
of
\begin{equation}
P_{\rho_{\Psi_{2,4,s}}\,\pm \pm \pm \pm} ({\hat \theta},{\hat \varphi} )=
{\rm Tr}\left[\rho_{\Psi_{2,4,s}} \cdot F_{\pm \pm \pm \pm} \left({\hat \theta},{\hat \varphi} \right)\right].
\end{equation}

The expectation functions and joint probabilities to find the four particles
in an even or in an odd number of
spin~``$-$''-states when measured along
($\theta_1$, $\varphi_1 $),
($\theta_2$, $\varphi_2 $),
($\theta_3$, $\varphi_3 $),  and
($\theta_4$, $\varphi_4 $)
are enumerated in Table~\ref{2005-gtq-2part}.
%
%
In the following, omitted arguments are zero.
For example, the expectation function of the general singlet state
in Eq.~(\ref{2005-hp-ep24smgf})
restricted to
$\varphi_1=\varphi_2=\varphi_3=\varphi_4=0$
is
\begin{equation}
\begin{array}{lll}
E(\tau  ;{\hat \theta} )  &=&
\frac{1}{3}\Bigl(
\left[ 2 + \cos (2\,\tau  ) \right] \,\cos (\theta_1  - \theta_2 )\,\cos (\theta_3  - \theta_4 )\\
&&{}+ 2\,\sin \tau
         \left[\sin \tau  \,\cos (\theta_1  + \theta_2  - \theta_3  - \theta_4 ) +
       {\sqrt{3}}\,\cos \tau  \,\sin (\theta_1  - \theta_2 )\,\sin (\theta_3  - \theta_4 ) \right]
\Bigr)
\end{array}
  \label{2005-gtq-Expmgf}
\end{equation}
For $\tau  = 0$ and $\tau = \frac{\pi}{2}$, Eq.~(\ref{2005-gtq-Expmgf})
reduces to
$E_{\rho_{\Psi_{2,4,2}}}$
and
$E_{\rho_{\Psi_{2,4,1}}}$
in Table~\ref{2005-gtq-2part}, respectively.
\begin{table}
\begin{tabular}{c}
\hline\hline
Two-partite singlet state\\
$
P_==
{1\over2}\left(1 + E  \right)
\; ,\;
P_{\neq} =
{1\over2}\left(1 - E \right)
$
\\
$
E(\theta_1,\theta_2,\varphi_1 , \varphi_2)=P_= -P_{\neq }=
-\left[\cos \theta_1 \cos \theta_2 + \cos (\varphi_1 - \varphi_2) \sin \theta_1 \sin \theta_2\right]
$
\\
$
E(\theta_1,\theta_2)= -\cos (\theta_1 - \theta_2)
$
\\
$
E(\frac{\pi}{2},\frac{\pi}{2},\varphi_1 , \varphi_2) = - \cos (\varphi_1 - \varphi_2)
$
\\
\hline
Four-partite singlet states\\
$
P_{ \rm even} =
{1\over2}\left[1 + E  \right]
\; ,\;
P_{\rm odd} =
{1\over2}\left[1 - E  \right]
\; ,\;
E=
P_{{\rm even}}
-
P_{{\rm odd}}
$
\\
$
\begin{array}{lll}
E_{\rho_{\Psi_{2,4,1}}}({\hat \theta} ,{\hat \varphi})  &=&
\frac{1}{3}
\left\{
\cos \theta_3 \sin \theta_1
\left[
-\cos \theta_4 \cos (\varphi_1 - \varphi_2) \sin \theta_2 +
          2 \cos \theta_2 \cos (\varphi_1 - \varphi_4) \sin \theta_4
\right] +
\right.
\\
&&\qquad
    \sin \theta_1 \sin \theta_3
\left[2 \cos \theta_2 \cos \theta_4 \cos (\varphi_1 - \varphi_3)  +
\right.
\\
&&\qquad
\qquad
\left.
\left(
2 \cos (\varphi_1 + \varphi_2 - \varphi_3 - \varphi_4) +
                \cos (\varphi_1 - \varphi_2)
                \cos (\varphi_3 - \varphi_4)
\right) \sin \theta_2 \sin \theta_4
\right]   +
\\
&&\qquad
    \cos \theta_1
\left[
2 \sin \theta_2
\left(
\cos \theta_4 \cos (\varphi_2 - \varphi_3) \sin \theta_3 +
                \cos \theta_3 \cos (\varphi_2 - \varphi_4) \sin \theta_4
\right) \right.
 +
\\
&&\qquad
\qquad
\left.
\left.
\cos \theta_2
\left(3 \cos \theta_3 \cos \theta_4 -
                \cos (\varphi_3 - \varphi_4) \sin \theta_3
\sin \theta_4
\right)
\right]
\right\}
\end{array}
$
\\
$
E_{\rho_{\Psi_{2,4,1}}}( \frac{\pi}{2},\frac{\pi}{2},\frac{\pi}{2},\frac{\pi}{2},\hat \varphi )=
\frac{1}{3} \left[2 \cos (\varphi_1+\varphi_2- \varphi_3 - \varphi_4)
+\cos (\varphi_1-\varphi_2) \cos (\varphi_3-\varphi_4)
\right]
$ \\
$
E_{\rho_{\Psi_{2,4,1}}}({\hat \theta} )  =
\frac{1}{3} \left[2 \cos (\theta_1 +\theta_2 -\theta_3 -\theta_4 )+\cos
   (\theta_1 -\theta_2 ) \cos (\theta_3 -\theta_4 )\right]
$
\\
$E_{\rho_{\Psi_{2,4,2}}}({\hat \theta} )  =
\cos (\theta_1 -\theta_2 ) \cos (\theta_3 -\theta_4 )
$
\\
$
\begin{array}{lll}
E_{\rho_{\Psi_{2,4,2}}}({\hat \theta} , {\hat \varphi } )  &=&
\left[\cos \theta_1 \cos \theta_2 +
          \cos ( \varphi_1 - \varphi_2) \sin \theta_1 \sin \theta_2\right]\cdot \\
&&\qquad  \qquad  \left[\cos \theta_3 \cos \theta_4 +
          \cos (\varphi_3 - \varphi_4) \sin \theta_3 \sin \theta_4
\right]
\end{array}
$
\\
$
\begin{array}{lll}
E(\tau  ;{\hat \theta} )  &=&
\frac{1}{3}\left\{
\left[ 2 + \cos (2\,\tau  ) \right] \,\cos (\theta_1  - \theta_2 )\,\cos (\theta_3  - \theta_4 )+
 \right.  \\
&& \qquad \; \left. + 2\,\sin \tau
         \left[\sin \tau  \,\cos (\theta_1  + \theta_2  - \theta_3  - \theta_4 ) +
       {\sqrt{3}}\,\cos \tau  \,\sin (\theta_1  - \theta_2 )\,\sin (\theta_3  - \theta_4 ) \right]
\right\}
\end{array}
$
\\
$
\begin{array}{lll}
E(\tau  ;{\hat \theta},{\hat \varphi} )  &=&
\frac{1}{3} \left\{\cos \theta_1  \left(\cos \theta_2  \{3  \cos \theta_3  \cos \theta_4 +[2 \cos (2 \tau )+1]  \cos (\varphi_3-\varphi_4) \sin    \theta_3  \sin \theta_4 \}+ \right. \right. \\
&& \qquad  \qquad
2 \sin \theta_2  \sin \tau  \left[\cos \theta_3  \cos (\varphi_2-\varphi_4) \sin \theta_4 \left(\sqrt{3} \cos \tau +\sin \tau \right)- \right. \\
&& \qquad  \qquad
\left. \left.
\cos \theta_4  \cos (\varphi_2-\varphi_3) \sin \theta_3  \left(\sqrt{3} \cos \tau -\sin \tau \right)\right]\right)+\\
&& \qquad
\sin \theta_1 \left(\cos \theta_3  \left\{\cos \theta_4  [2 \cos (2 \tau )+1] \cos (\varphi_1-\varphi_2) \sin \theta_2 + \right.\right.\\
&&   \qquad  \qquad
\left.
2 \cos \theta_2  \cos (\varphi_1-\varphi_4) \sin \theta_4  \sin \tau  \left(\sin \tau -\sqrt{3} \cos \tau \right)\right\}+\\
&&   \qquad  \qquad
\sin \theta_3  \left[2 \cos \theta_2  \cos \theta_4  \cos (\varphi_1-\varphi_3) \sin \tau  \left(\sqrt{3} \cos \tau +\sin \tau \right)+\right.\\
&&   \qquad  \qquad
\sin \theta_2  \sin \theta_4  \left\{2 \cos (\varphi_1+\varphi_2-\varphi_3-\varphi_4) \sin ^2\tau +\right.\\
&&  \qquad  \qquad
\left.  \left. \left. \left.
[\cos (2 \tau )+2] \cos (\varphi_1-\varphi_2)\cos (\varphi_3-\varphi_4)+\sqrt{3} \sin (2 \tau ) \sin (\varphi_1-\varphi_2) \sin (\varphi_3-\varphi_4)
\right\}\right]\right)\right\}
\end{array}
$
\\
\hline\hline
\end{tabular}
\caption{Probabilities and expectation functions
for finding an odd or even number of spin~``$-$''-states.
Omitted arguments are zero.
\label{2005-gtq-2part}
}
\end{table}

We concentrate on the algebraic evaluation of  $E_{\rho_{\Psi_{2,4,1}}}$,
as this expectation function is from a nontrivial
non-zigzag singlet state and thus can be expected to reveal additional structure not inherited from the two-partite correlations
also enumerated in Table~\ref{2005-gtq-2part}.
If all the polar angles $\hat \theta$ are all set to $\pi /2$,
then this correlation function yields
\begin{equation}
E_{\rho_{\Psi_{2,4,1}}}( \frac{\pi}{2},\frac{\pi}{2},\frac{\pi}{2},\frac{\pi}{2},\hat \varphi )=
\frac{1}{3} \left[2 \cos (\varphi_1+\varphi_2- \varphi_3 - \varphi_4)
+\cos (\varphi_1-\varphi_2) \cos (\varphi_3-\varphi_4)
\right]
.
\end{equation}
Likewise, if all the azimuthal angles $\hat \varphi$ are all set to zero, one obtains
\begin{equation}
E_{\rho_{\Psi_{2,4,1}}}(\hat \theta )=
\frac{1}{3} \left[2 \cos (\theta_1+\theta_2- \theta_3 - \theta_4)
+\cos (\theta_1-\theta_2) \cos (\theta_3-\theta_4)
\right]
.
\end{equation}

\subsection{Plasticity of expectation function}

\begin{figure}[htbp]
  \centering
\begin{tabular}{cc}
  \includegraphics[width=60mm]{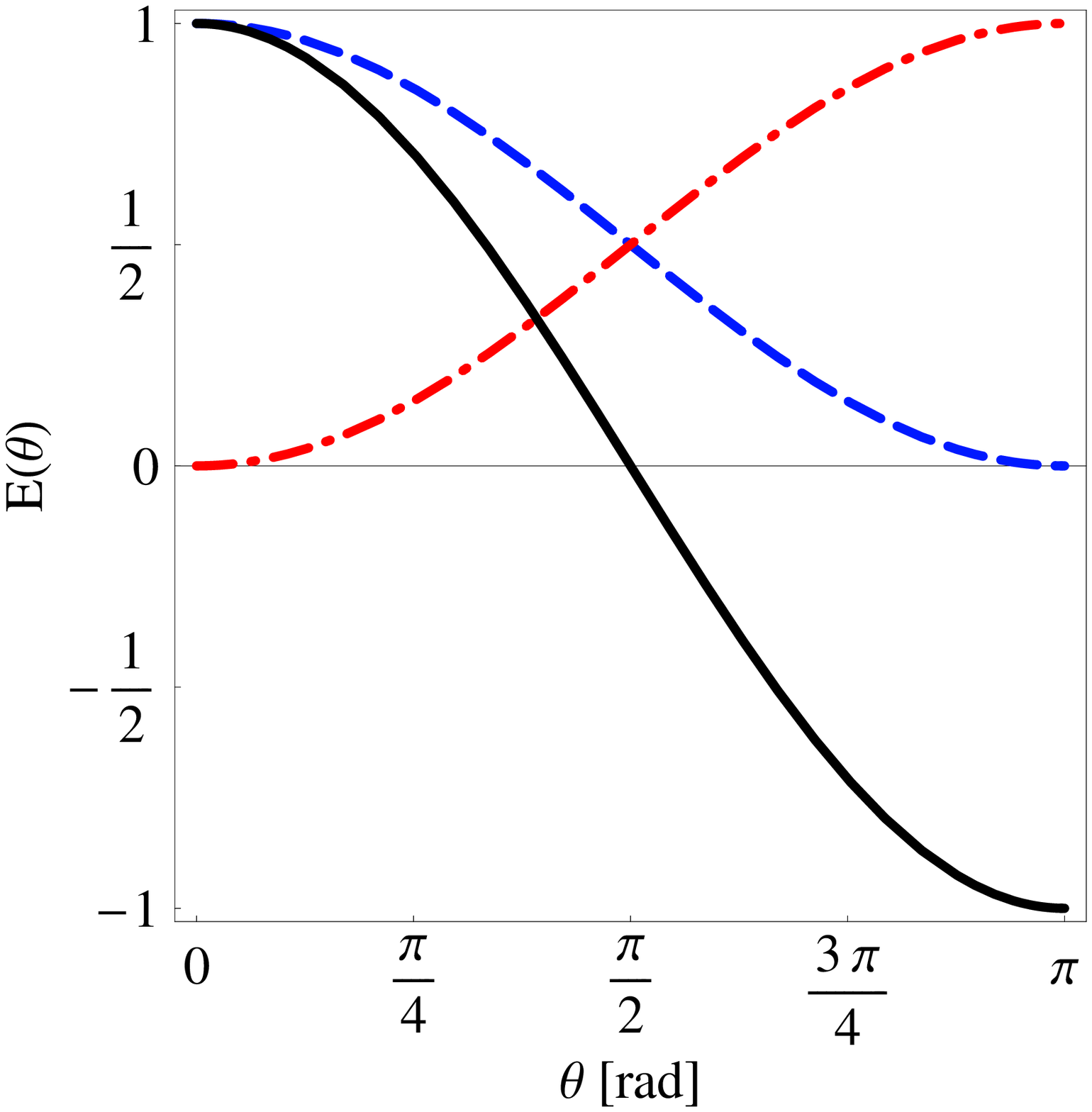}&
  \includegraphics[width=60mm]{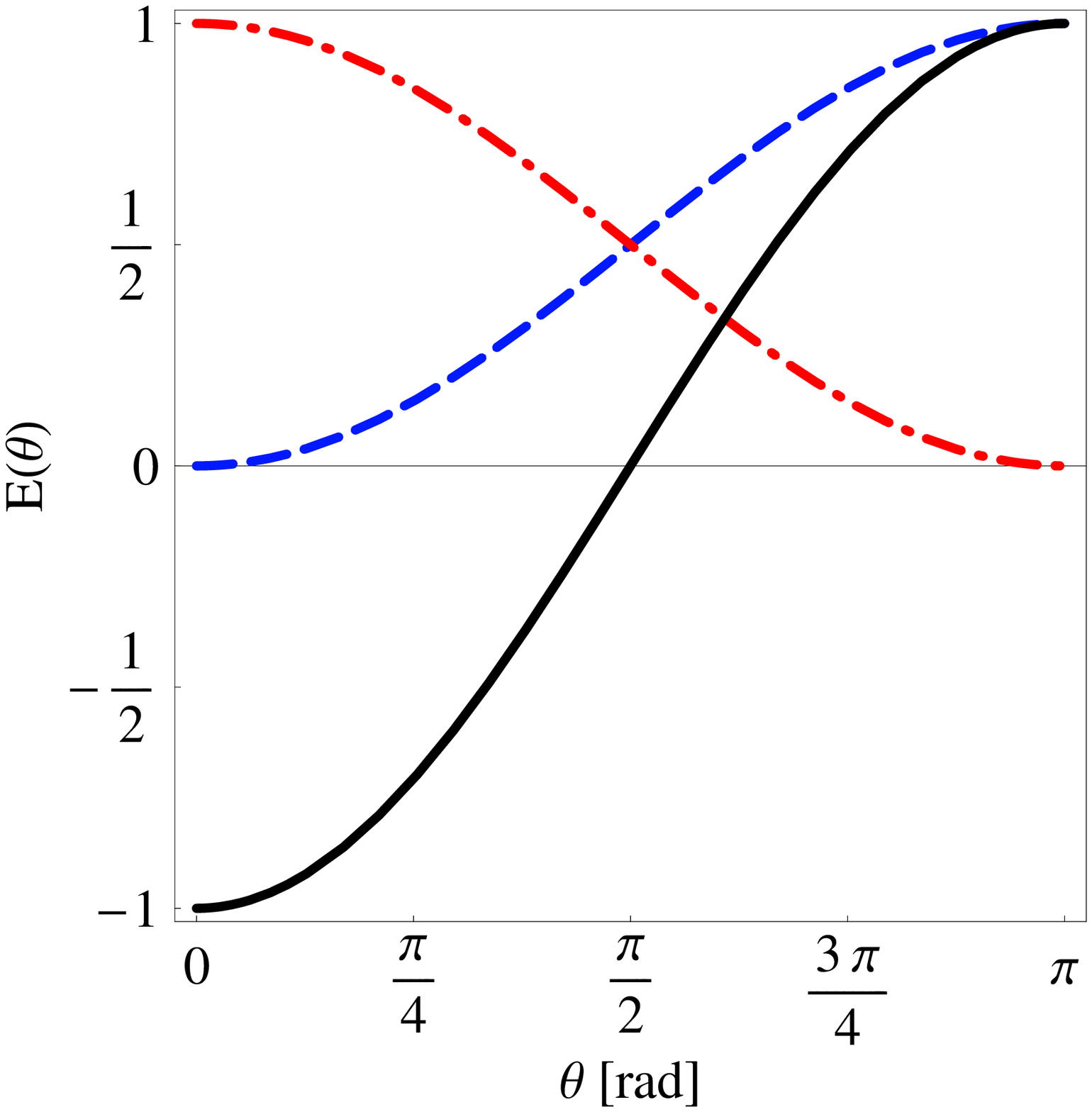}\\
\quad \qquad (a) &\quad  \quad (b)\\
  \includegraphics[width=60mm]{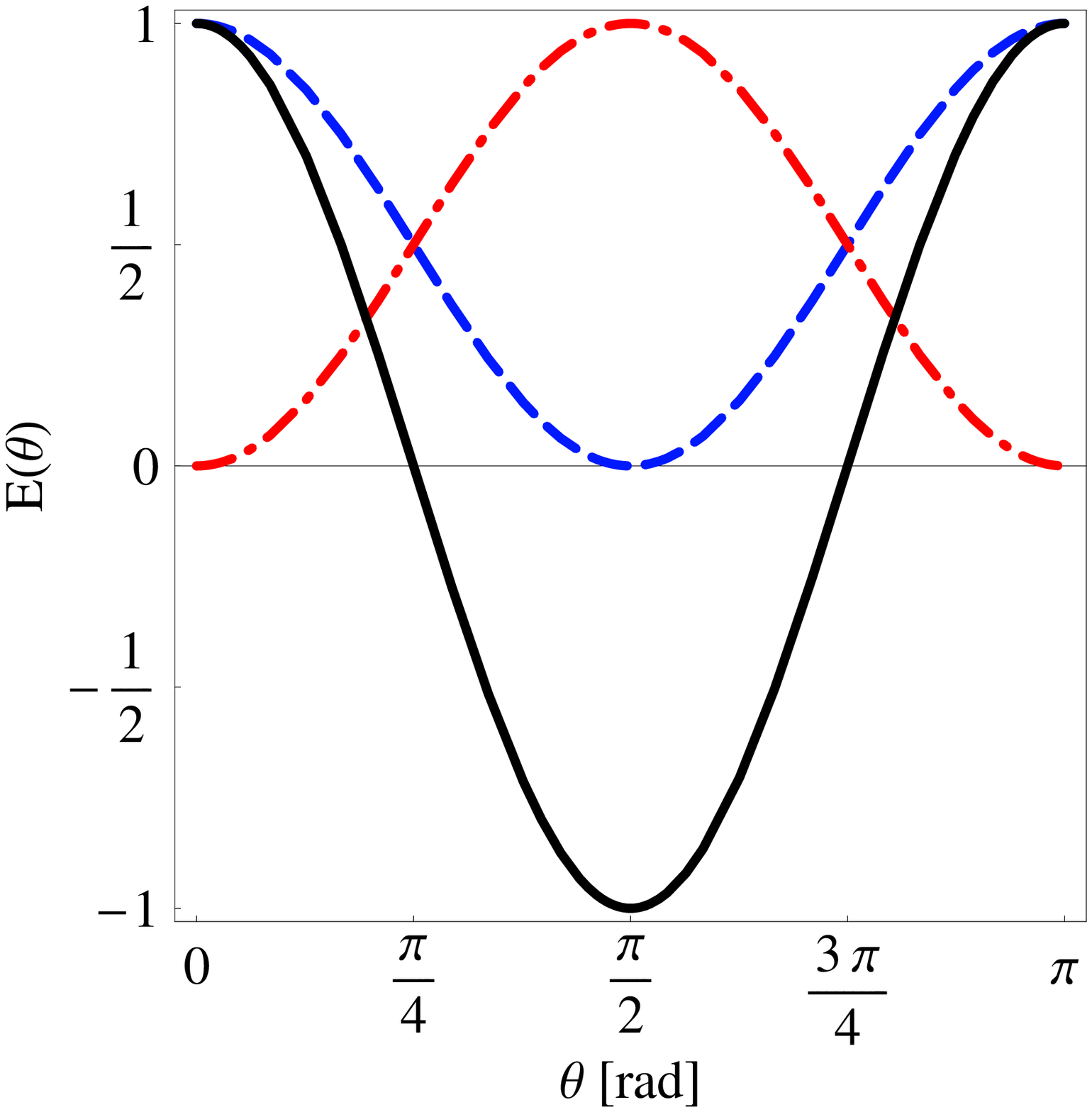}&
  \includegraphics[width=60mm]{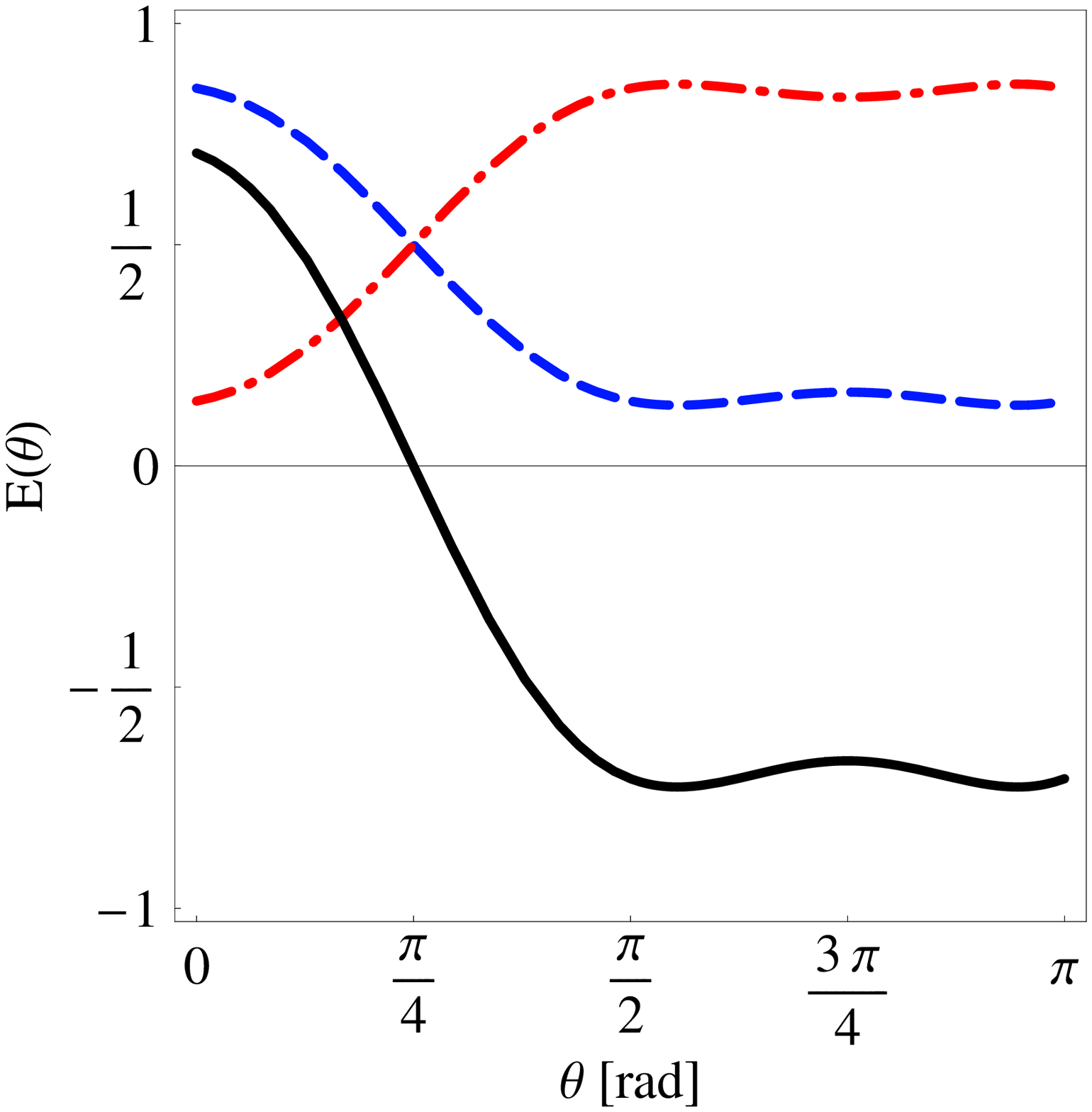}\\
\quad  \qquad (c)& \quad  \qquad (d)\\
  \includegraphics[width=60mm]{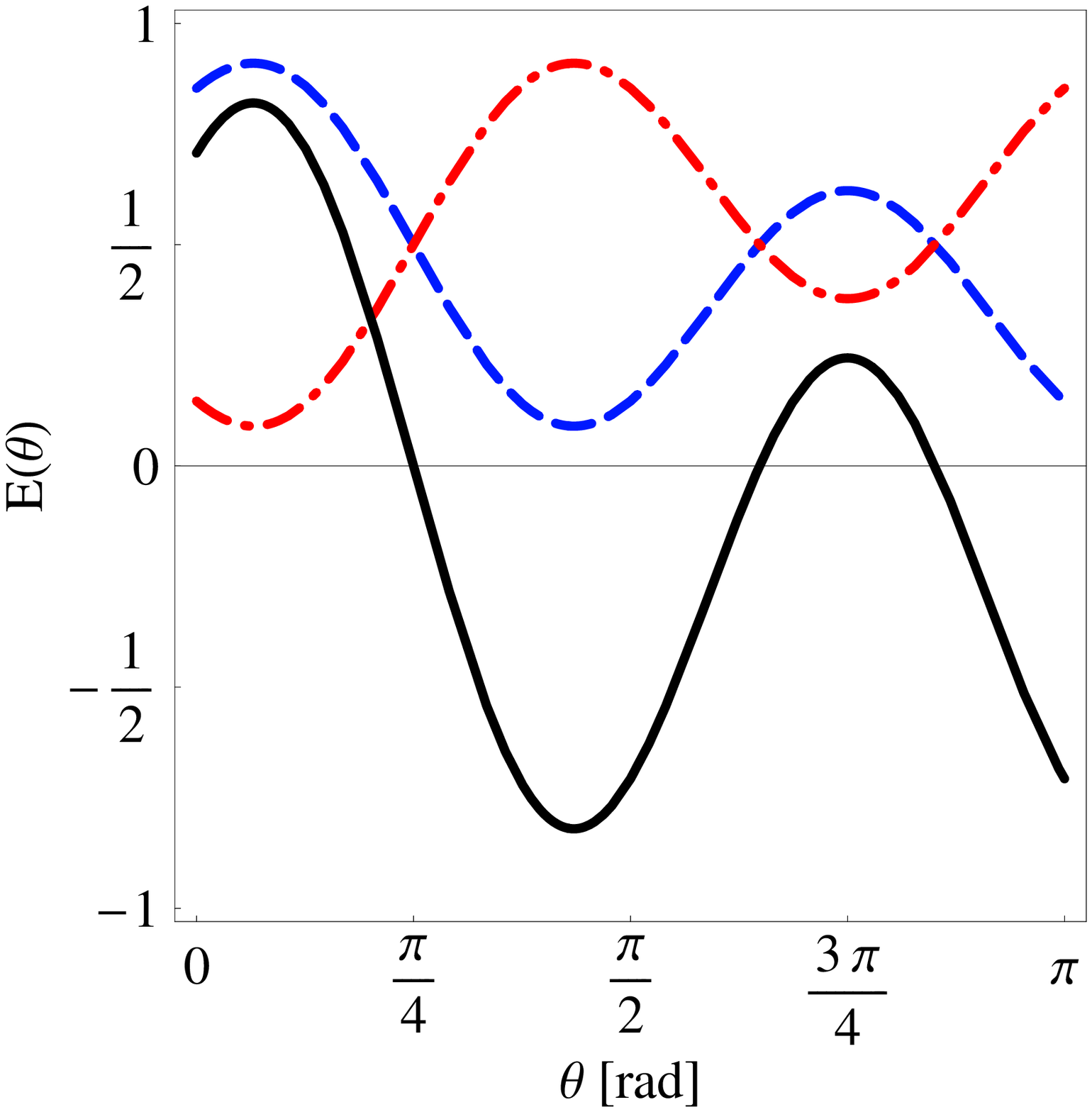}&
  \includegraphics[width=60mm]{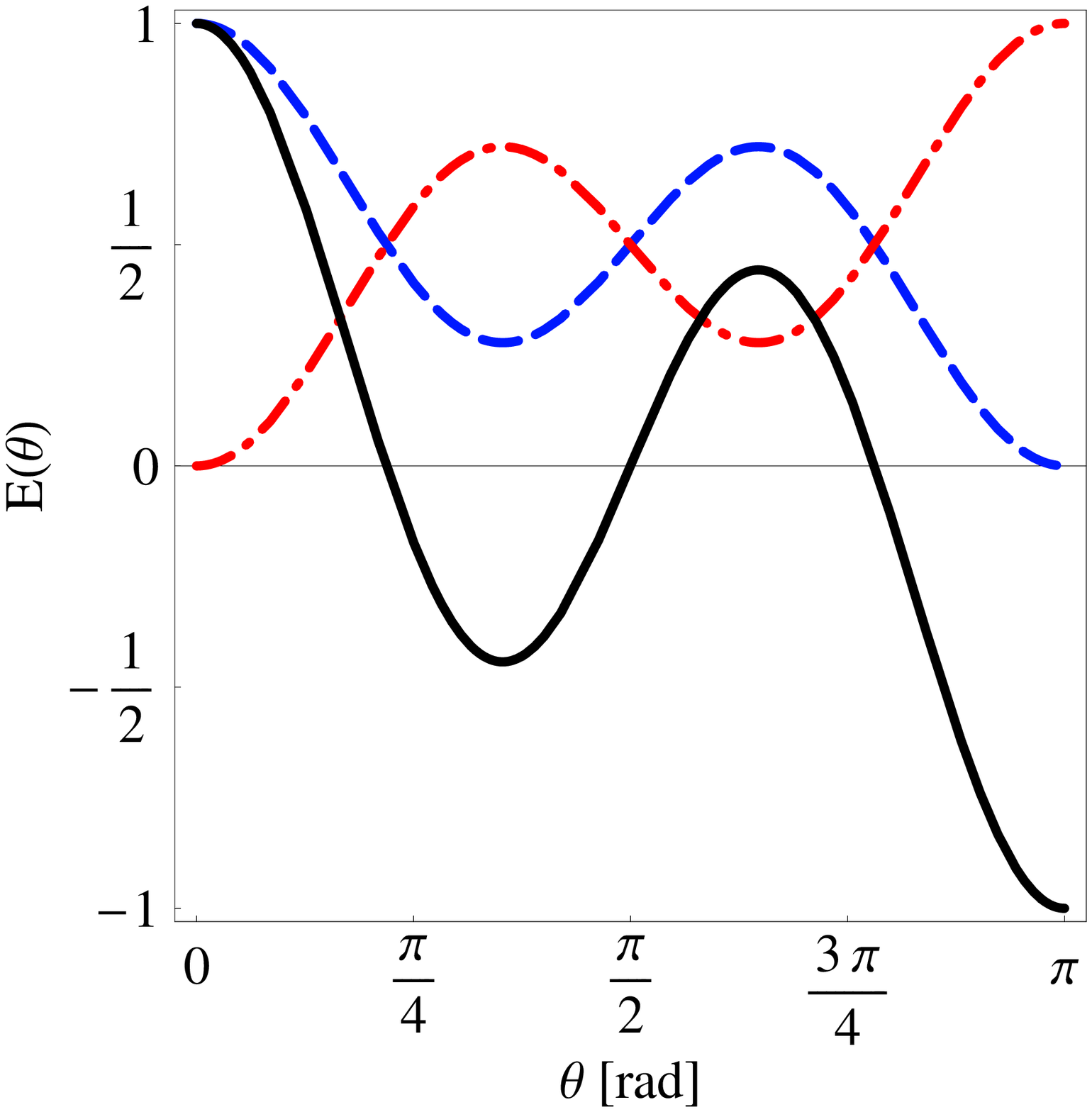}\\
\quad  \qquad (e)& \quad  \qquad (f)\\
\end{tabular}
  \caption{Probabilities and expectation values for
(a)
$\tau =0$,
$\theta_1 =\theta$,
$\theta_2 =\theta_3 =\theta_4 =0$,
(b)
$\tau =0$,
$\theta_1 =\theta$,
$\theta_2 =\theta_3  =0$,
$\theta_4 =\pi$,
(c)
$\tau =\frac{\pi}{2}$,
$\theta_1 =\theta_2 =-\theta_3 =\theta_4  =\theta$,
(d)
$\tau =\frac{\pi}{2}$,
$\theta_1 =-\theta_3 =\theta_4  =\theta$,
$\theta_2 =\frac{\pi}{4}$,
(e)
$\tau =\frac{\pi}{4}$,
$\theta_1 =-\theta_3 =\theta_4  =\theta$,
$\theta_2 =\frac{\pi}{4}$,
(f)
$\tau =\frac{\pi}{4}$,
$\theta_1 =- \theta_3=\theta_4 =\theta$,
$ \theta_2 =0$.
Dashed (dash dotted) lines indicate probabilities to find an even (odd)
number of ``$-$'' outcomes, solid lines depict expectation functions.
}
\label{2005-gtq-fpe}
\end{figure}

The plasticity of the expectation function
$E(\tau  ;{\hat \theta} )$ is comparable to the two-particle
expectation function $E(\theta)=-\cos \theta$
for measurements in one plane can  be demonstrated by plotting
the probabilities and expectation values for selectively chosen parameters,
as depicted in Fig.~\ref{2005-gtq-fpe}.

As there are four particles involved, the outcomes of one or two particles can be utilized
to select the events of the other particles.
Let ``$\pm_i$'' stand for the observation of spin state plus or minus on the $i$th particle.
Table~\ref{2005-gtq-2partwith}
contains the results of the associated expectation values and joint probabilities
for finding an odd or even number of spin~``$-$''-states.
\begin{table}
\begin{tabular}{c}
\hline\hline
Three-partite GHZM state (Ref.~\cite{krenn1})\\
$
P_\pm  =
{1\over 4}\left[1 + 2 E  \right]
\; ,\;
P_\pm  =
{1\over 4}\left[1 - 2 E  \right]
\; ,\;
E_\pm =
P_\pm
-
P_\pm
$
\\
$E_\pm (\theta_1,\theta_2,\theta_3,\varphi_1, \varphi_2,\varphi_3 \vert \pm_3)  =
\frac{1}{2}\left[
\cos \theta_1 \cos \theta_2
  \pm_3 \cos \left(\varphi_1 + \varphi_2 + \varphi_3\right)
\sin \theta_1 \sin \theta_2 \sin \theta_3
\right]$\\
\hline
Four-partite singlet states\\
$
E_{\rho_{\Psi_{2,4,1}}}({\hat \theta} ,{\hat \varphi}\vert \pm_4)=   \frac{1}{12}
\pm {1\over 2} E_{\rho_{\Psi_{2,4,1}}}({\hat \theta} ,{\hat \varphi})
$   \\
$
E_{\rho_{\Psi_{2,4,1}}}({\hat \theta} \vert \pm_3 \pm_4)=   \frac{1}{12}\left\{2 (\pm_3 1) (\pm_4 1) \cos (\theta_1 + \theta_2 - \theta_3 - \theta_4) +
    \cos (\theta_1 - \theta_2) \left[1 + (\pm_3 1) (\pm_4 1) \cos (\theta_3 - \theta_4)\right]\right\}
$   \\
$
\begin{array}{lll}
E_{\rho_{\Psi_{2,4,1}}}({\hat \theta} ,{\hat \varphi}\vert \pm_3 \pm_4)&=&  \frac{1}{12}
\left\{\cos \theta_1 (2 (\pm_3 1) (\pm_4 1) \sin \theta_2 \left[\cos \theta_4 \cos (\varphi_2 - \varphi_3) \sin \theta_3 + \right. \right.
\\ && \qquad
\left.
                \cos \theta_3 \cos (\varphi_2 - \varphi_4) \sin \theta_4\right] +
\\ && \qquad
                \cos \theta_2 \left[1 + 3 (\pm_3 1) (\pm_4 1) \cos \theta_3 \cos \theta_4 -         \right.
\\ && \qquad
\left.
                (\pm_3 1) (\pm_4 1) \cos (\varphi_3 - \varphi_4) \sin \theta_3 \sin \theta_4\right]) +
\\ && \qquad
                \sin \theta_1 (\cos (\varphi_1 - \varphi_2) \sin \theta_2 \left[1 - (\pm_3 1) (\pm_4 1) \cos \theta_3 \cos \theta_4 +\right.
\\ && \qquad
\left.
                (\pm_3 1) (\pm_4 1) \cos (\varphi_3 - \varphi_4) \sin \theta_3 \sin \theta_4\right] +
\\ && \qquad
                2 (\pm_3 1) (\pm_4 1) \left[\cos \theta_2 \cos \theta_4 \cos (\varphi_1 - \varphi_3) \sin \theta_3 +\right.
\\ && \qquad
                \cos \theta_2 \cos \theta_3 \cos (\varphi_1 - \varphi_4) \sin \theta_4 +
\\ && \qquad
\left.
\left.
\left.
                \cos (\varphi_1 + \varphi_2 - \varphi_3 - \varphi_4) \sin \theta_2 \sin \theta_3 \sin \theta_4\right]\right)\right\}
\end{array}$   \\
$
E_{\rho_{\Psi_{2,4,1}}}({\hat \theta} \vert \pm_2 \pm_4)=   \frac{1}{12}\left\{
     (\pm_2 1) (\pm_4 1) \left[2 \cos (\theta_1 + \theta_2 - \theta_3 - \theta_4) +
          \cos (\theta_1 - \theta_2) \cos (\theta_3 - \theta_4)\right]
-2 \cos (\theta_1 - \theta_3)
\right\}
$   \\
$
\begin{array}{lll}
E_{\rho_{\Psi_{2,4,1}}}({\hat \theta} ,{\hat \varphi}\vert \pm_2 \pm_4)&=&  \frac{1}{12}
\left\{\cos \theta_1 ((\pm_2 1) (\pm_4 1) \sin \theta_3 \left[2 \cos \theta_4 \cos (\varphi_2 - \varphi_3) \sin \theta_2 -  \right.\right.
\\ && \qquad
\left.
                \cos \theta_2 \cos (\varphi_3 - \varphi_4) \sin \theta_4\right] +
\\ && \qquad
           \cos \theta_3 \left[-2 + 3 (\pm_2 1) (\pm_4 1) \cos \theta_2 \cos \theta_4 + \right.
\\ && \qquad
\left.
                2 (\pm_2 1) (\pm_4 1) \cos (\varphi_2 - \varphi_4) \sin \theta_2 \sin \theta_4\right]) +
\\ && \qquad
    \sin \theta_1 ((\pm_2 1) (\pm_4 1) \cos \theta_3 \left[-\cos \theta_4 \cos (\varphi_1 - \varphi_2) \sin \theta_2 +\right.
\\ && \qquad
\left.
                2 \cos \theta_2 \cos (\varphi_1 - \varphi_4) \sin \theta_4\right] +
\\ && \qquad
          \sin \theta_3 (2 \left[-1 + (\pm_2 1) (\pm_4 1) \cos \theta_2 \cos \theta_4\right] \cos (\varphi_1 - \varphi_3) +
\\ && \qquad
                (\pm_2 1) (\pm_4 1) (2 \cos (\varphi_1 + \varphi_2 - \varphi_3 - \varphi_4) +
\\ && \qquad
\left.
                      \cos (\varphi_1 - \varphi_2) \cos (\varphi_3 - \varphi_4)) \sin \theta_2 \sin \theta_4))\right\}
\end{array}$   \\
\hline\hline
\end{tabular}
\caption{Probabilities and expectation functions
for finding an odd or even number of spin~``$-$''-states
with selection. ``$\pm_i$'' stands for the observation of spin state plus or minus on the $i$th particle.
\label{2005-gtq-2partwith}
}
\end{table}

Two or three observables could also be grouped together to form a ``condensed'' observable.
For instance, for each individual quadruple of outcomes $\{o_1,o_2,o_3,o_4\}$
the values of the first and the
second, as well as of the third and the
fourth particle could be multiplied to obtain two other, dichotomic observables
$o_1o_2$ and $o_3o_4$, respectively. More generally, one could take all partitions of 4,
such that the outcomes of all particles within an element of the partition
are multiplied. As the single outcomes occur at random, their resulting products
and thus the new condensed observable would also represent random variables. Since the multiplication is associative,
the resulting condensed correlations are just the four-partite correlations discussed
so far.

\section{Summary}

In summary, we have discussed an algorithmic procedure to enumerate all singlet states of $N$ particles of arbitrary spin.
We have then explicitly enumerated the first cases for spin one-half and spin one and discussed their symmetries.
These results have then be applied for a calculation of the quantum probabilities and expectation functions of
four spin one-half particles in four arbitrary directions.
We conclude by pointing out that all discussed configurations could,
as a proof of principle, be locally realized
by generalized beam splitters \cite{rzbb,zukowski-97,svozil-2004-analog}.

{\bf Dedication and acknowledgments:}
This paper is dedicated to Sylvia Pulmannova, in appreciation of her scientific accomplishments,
and also for the enjoyment and privilege to be co-author in some of her research articles, as well as meeting her personally.
One of the most remarkable impressions on me (K.S.) was her humbleness as a person and colleague, accompanied by her extreme professional skills in mathematics, logics and physics.
The authors thank Rainer Dirl and Peter Kasperkovitz for their advise in group theoretical questions; K.S. gratefully acknowledges discussions with Boris Kamenik, G\"unther Krenn and Johann Summhammer in Viennese coffee houses and elsewhere.


\end{document}